\RequirePackage{fix-cm}
\pdfoutput=1
\documentclass[twocolumn, final]{svjour3}          
\usepackage{graphicx}
\usepackage{bbm}
\usepackage{latexsym}
\usepackage{amssymb,amsmath}
\usepackage{hyperref}
\newcommand{\ohm}{$\mathrm{\Omega}$}
\journalname{..}

\begin{document}

\title{Synchronization in hyperchaotic time-delayed electronic oscillators coupled indirectly via a common environment}

\titlerunning{Synchronization in hyperchaotic time-delayed electronic oscillators coupled indirectly via a common environment}        

\author{Tanmoy Banerjee \and Debabrata Biswas}


\institute{T. Banerjee$^{*}$ \and D. Biswas \at
              Department of Physics, The University of Burdwan, Burdwan 713 104, West Bengal, India.\\
              $^*$\email{tanbanrs@yahoo.co.in},~{tanban.buphys@gmail.com}}

\date{Received: date / Accepted: date}

\maketitle

\begin{abstract}
The present paper explores the synchronization scenario of hyperchaotic time-delayed electronic oscillators coupled indirectly via a common environment. We show that depending upon the coupling parameters a hyperchaotic time-delayed system can show in-phase or complete synchronization, and also inverse-phase or anti-synchronization. This paper reports the {\it first} experimental confirmation of synchronization of hyperchaos in time-delayed electronic oscillators coupled indirectly through a common environment. We confirm the occurrence of in-phase and inverse-phase synchronization phenomena in the coupled system through the dynamical measures like generalized autocorrelation function, correlation of probability of recurrence, and the concept of localized sets computed directly from the experimental time-series data. We also present a linear stability analysis of the coupled system. The experimental and analytical results are further supported by the detailed numerical analysis of the coupled system. Apart from the above mentioned measures, we numerically compute another quantitative measure, namely, Lyapunov exponent spectrum of the coupled system that confirms the transition from the in-phase (inverse-phase) synchronized state to the complete (anti-) synchronized state with the increasing coupling strength. 
\end{abstract}
\section{Introduction}
\label{intro}
For more than two decades, synchronization of chaos and hyperchaos have been an active field of research in various fields, including  physics, biology, mathematics, engineering, etc. After the seminal paper of Pecora and Carrol \cite{pc} on chaos synchronization, several synchronization schemes and processes have been observed and identified; example inludes, complete  synchronization,  generalized synchronization, phase synchronization, lag synchronization, anticipatory synchronization , impulsive synchronization, etc. (for detailed discussion refer to two excellent review papers \cite{pcrev,rev} and references therein). 

After the initial emphasis on the synchronization of low dimensional chaotic systems (i.e., chaotic systems having a single positive Lyapunov exponent), researchers soon attracted towards the study of synchronization in high-dimensional systems like complex network \cite{forprl}, and nonlinear delay dynamical system (DDS). Particularly, studies on the synchronization scenario of delay dynamical systems are challenging, both from theoretical and experimental point of view, owing to their infinite dimensionality  with a large number of positive Lyapunov exponents. Further, in real world most of the time we have to encounter with time-delay; few examples are, blood production in patients with leukemia (Mackey-Glass model) \cite{mac}, dynamics of optical systems (e.g. Ikeda system) \cite{ikeda1}\cite{ikeda2}, population dynamics \cite{pop}, physiological model \cite{tumer}, El Ni\~{n}o/southern oscillation (ENSO) \cite{nino}, neural networks \cite{nural}, control system \cite{{control1},{control2},{exnd}}, etc. The first study on synchronization of chaos in time-delayed system has been reported by Pyragas \cite{py}. Later, several genres of synchronizations in time-delay systems have been reported; few examples are (without claiming to be complete): Lag and anticipatory synchronization \cite{sahalag,sahaas,lak05pre,lak11,tanls}, complete and generalized synchronization \cite{zhan,saha,tangs}, etc. The phenomenon of phase synchronization in time-delayed systems has first been reported by Senthilkumar et al \cite{lakps}  and later it was confirmed experimentally in Ref. \cite{lakpsexpt}. Recently, global phase synchronization and zero-lag synchronization (ZLS) in time-delay systems have been reported in \cite{gps} and \cite{zls}, respectively, showing the ongoing interest in the field of synchronization of time-delay systems. In all the above mentioned works (except \cite{zls}), the coupling scheme is essentially the {\it direct coupling}, i.e., either unidirectional coupling or bi-directional coupling, where either of the two coupled systems or both the systems directly affect the dynamics of  each other. Ref. \cite{zls} considers mutual and subsystem coupling configurations via dynamical relaying and reports the experimental confirmation of ZLS in a system of three coupled piecewise linear time-delay circuits. 

An important coupling scheme has recently been reported, namely the {\it environmental coupling} , which is an {\it indirect coupling} \cite{kat} scheme. Here two or more systems interact indirectly with each other through a common environment. The environment makes the systems synchronous without affecting their central dynamical features. Environmental coupling is particularly important in biological systems, e.g., populations of cells in which oscillatory reactions are taking place interact with each other via chemicals that diffuse in the surrounding medium \cite{kat}. The environmental coupling scheme has been investigated numerically in detail by Reshmi et al \cite{ambika} where the phase and complete (anti-) synchronization of environmentally coupled low-dimensional chaotic systems (R\"{o}ssler and Lorenz systems) are explored using Lyapunov exponent spectrum, average phase errors, and correlation functions. A linear stability analysis has also been reported in that paper. Later, the experimental study of the same has been reported in \cite{srimali}. Amplitude death in environmentally coupled chaotic systems are reported in \cite{{ambika1},{srimali1}}. In this context, another important work has recently been reported by Sharma et al. \cite{dana} that considers a variant of environmental coupling to show (both numerically and experimentally) the occurrence of phase-flip bifurcation in periodic and chaotic systems.

All the previous works on environmental coupling are restricted to the study of low-dimensional chaotic systems. The same for the time-delayed hyperchaotic system has not been reported yet; particularly experimental realization of environmentally coupled hyperchaotic time-delayed system is challenging enough, and, to the best of our knowledge, yet to be reported.  In this paper we study the synchronization states of two hyperchaotic time-delay systems that are coupled indirectly via a common environment. For the present study we choose a hyperchaotic time-delay system recently proposed in \cite{banerjee12}. The choice of this system is led by the fact that unlike other time-delay systems it poses three important features that are useful for the experimental design: first, it has a closed form mathematical function for the the nonlinearity (unlike piece-wise-linear nonlinearity), second, it shows hyperchaos even for a small or moderate value of time-delay, and lastly, and most importantly, it can be realized with off-the-shelf electronic circuit components. Further, the hyperchaotic attractor of the system is inherently phase-incoherent, thus the present study can be extended to other general class of hyperchaotic systems. We show that depending upon the coupling parameters a hyperchaotic time-delayed system can show in-phase or complete synchronization, and also inverse-phase or anti-synchronization. For the {\it first} time we report the experimental studies of synchronization of hyperchaos in time-delayed electronic oscillators coupled indirectly through a common environment. We confirm the occurrence of phase synchronization in the coupled system through the dynamical measures like generalized autocorrelation function, correlation of probability of recurrence \cite{{cpr1},{cpr2}}, and the concept of localized sets \cite{kurth_local} computed directly from the experimental time-series data. We also perform a linear stability analysis of the coupled system for the complete and anti-synchronized cases. The experimental results are further supported by the detailed numerical simulations of the coupled system. Numerical computations are carried out to find out the Lyapunov exponent spectrum of the coupled system that confirms the transition from the in-phase (inverse-phase) synchronized state to the complete (anti-) synchronized state with the increasing coupling strength. Numerical recurrence analysis and concept of localized set are used to reconfirm the occurrence of phase sysnchronization in the coupled system.

The paper is organized in the following manner: the next section describes the environmental coupling scheme for a time-delay system, and also describes the system under consideration and a summary of its main dynamical behavior. Section \ref{sec:3} describes the experimental design of the coupled system. Experimental results, recurrence analysis and other dynamical measures are presented in this section. Sect.\ref{sec:4} gives an account of linear stability analysis of the synchronized state. Sect.~\ref{sec:5} describes the numerical simulation results of the coupled system, Lyapunov exponent spectrum, recurrence analysis, concept of localized set, and two-parameter stable zone of synchronization have been reported in this section. Finally, Sect. \ref{sec:6} concludes the outcome and importance of the whole study.
\section{Environmentally coupled time-delayed system}
\label{sec:2}
\subsection{Environmental coupling}
Let us consider two first-order identical time-delayed systems coupled indirectly through a common environment $z$. The mathematical model of the coupled system is given by 
\begin{subequations}\label{dr}
\begin{align}
\dot{x}&=-ax(t)+b_1f(x_{\tau})+\epsilon_1\beta_1z,\label{x}\\
\dot{y}&=-ay(t)+b_2f(y_{\tau})+\epsilon_1\beta_2z,\label{y}\\
\dot{z}&=-\kappa z-\frac{\epsilon_2}{2}(\beta_1x+\beta_2y).\label{z}
\end{align}
\end{subequations}
where $a>0$, $b_1$ and $b_2$ are called the feedback rates for the $x$-system and $y$-system, respectively. Also, $u_{\tau} \equiv u(t-\tau)$ ($u=x,y$), where $\tau \in \mathbb{R}^{+}$ is a constant  time delay. $\epsilon_1$ determines the coupling strength between environment and the systems that controls the effect of environment on the systems. $\epsilon_2$ determines the coupling strength between the system and environment that controls the effect of individual systems on the environment. $\beta_1$ and $\beta_2$ determine the nature of coupling: for $\beta_1=1, \beta_2=-1$ the systems are attractively coupled; on the other hand repulsive coupling is achieved for $\beta_1=1$ and $\beta_2=1$.  Finally, $\kappa$ ($>0$) determines the nature of the environment; in absence of both the $x$- and $y$- systems (i.e. $\epsilon_2=0$), the environment decays towards the zero steady state and remains in that dormant state.      
\subsection{System description and dynamics of the uncoupled system}
\begin{figure}
 \includegraphics[width=.4\textwidth]{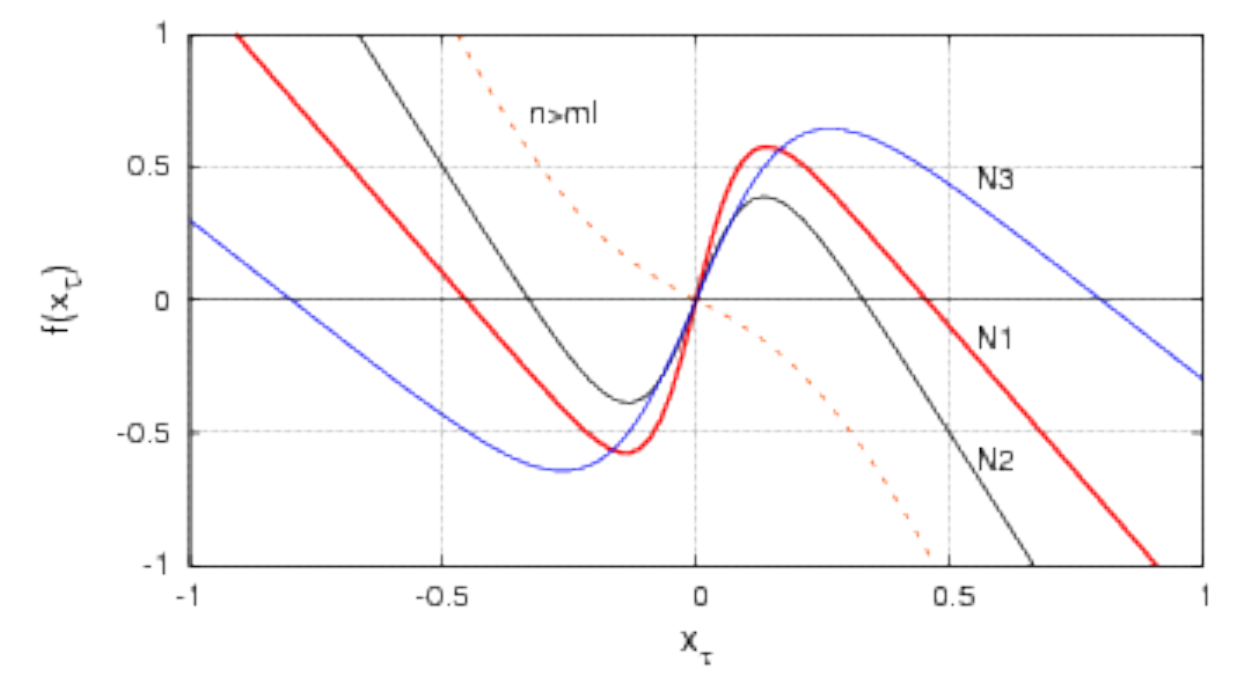}
 \caption{Nonlinearity with the function $f(x_{\tau})=-nx_{\tau}+m\tanh(lx_{\tau})$ with {\bf ``N1":} $n=2.2$, $m=1$, $l=10$; {\bf ``N2":} $n=3$, $m=1$, $l=8$, {\bf ``N3":} $n=1.5$, $m=1.2$, $l=8$. Dashed curve is for $n=4$, $m=1$, $l=3$, which shows that for $n>ml$ the nonlinearity does not show the two-humped nature.}
 \label{nl}
\end{figure}
\begin{figure}
  \includegraphics[width=.49\textwidth]{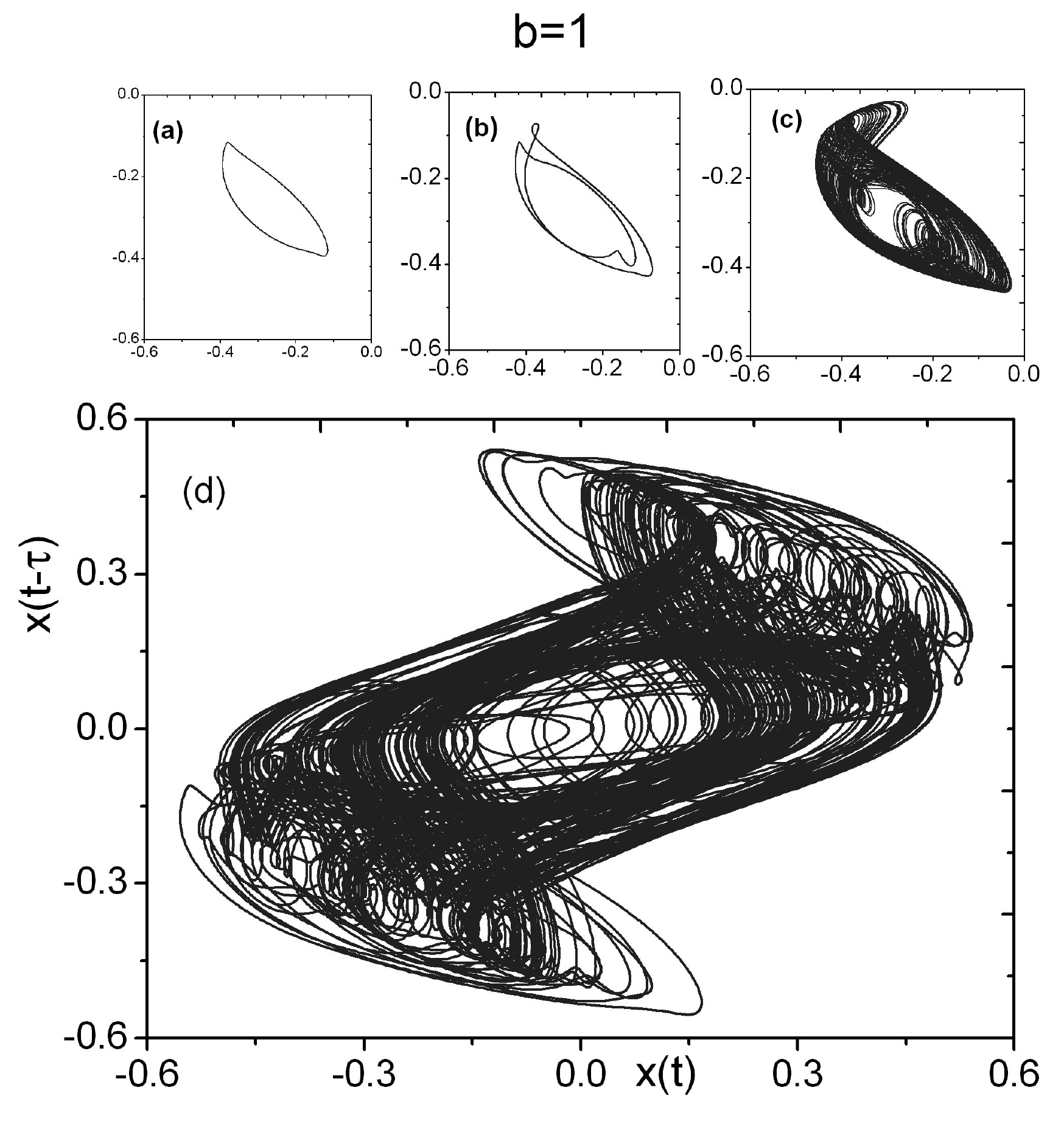}
\caption{Phase plane plot in $x$--$x(t-\tau)$ space for different $\tau$ ($b=1$): (a) $\tau=1.40$ (period-1), (b) $\tau=1.72$ (period-2), (c) $\tau=1.94$ (chaos), (d) $\tau=3.6$ (double scroll hyperchaos). (other parameters are: $a=1$, $n=2.2$, $m=1$, $l=10$).}
\label{phtau}       
\end{figure}
In this section we describe the time-delayed system proposed in Ref. \cite{banerjee12}, and also briefly discuss its important dynamical features. Ref. \cite{banerjee12} proposed the following first-order, nonlinear, retarded type delay differential equation with a single constant delay:
\begin{equation}\label{m1}
      \dot{x}(t) = -ax(t)+bf(x_\tau),
\end{equation}
where $a>0$ and $b$ are system parameters. In \cite{banerjee12} we consider both positive and negative values of $b$, but in this paper we will consider only the case $b>0$. 
Now, we define the following closed form mathematical function for the nonlinearity:
\begin{equation}\label{hw1} 
f(x_\tau) = -nx_{\tau}+m\tanh(lx_{\tau}),
\end{equation}
where $n$, $m$ and $l$ are all positive system parameters and they are restricted by the following constraint: $n<ml$. It can be seen that the nonlinear function is constituted by the weighted superposition of two functions, namely, the linear  proportionality function and the nonlinear $\tanh$ function. Further, $f(x_\tau)$ is an odd-symmetric function, i.e., $f(-x_\tau)=-f(x_\tau)$. 

Fig.\ref{nl} shows the nature of the nonlinearity produced by $f(x_\tau)$ for different values of $n$, $m$ and $l$. The nonlinearity shows a hump in the first quadrant and the third quadrant. The condition $n<ml$ ensures the two-humped nature of the nonlinear function. Also, \eqref{hw1} has another distinct advantage -- it provides a large number of choices of  $n$, $m$ and $l$ for which the two-humped nature will be preserved. 

\begin{figure}
  \includegraphics[width=.49\textwidth]{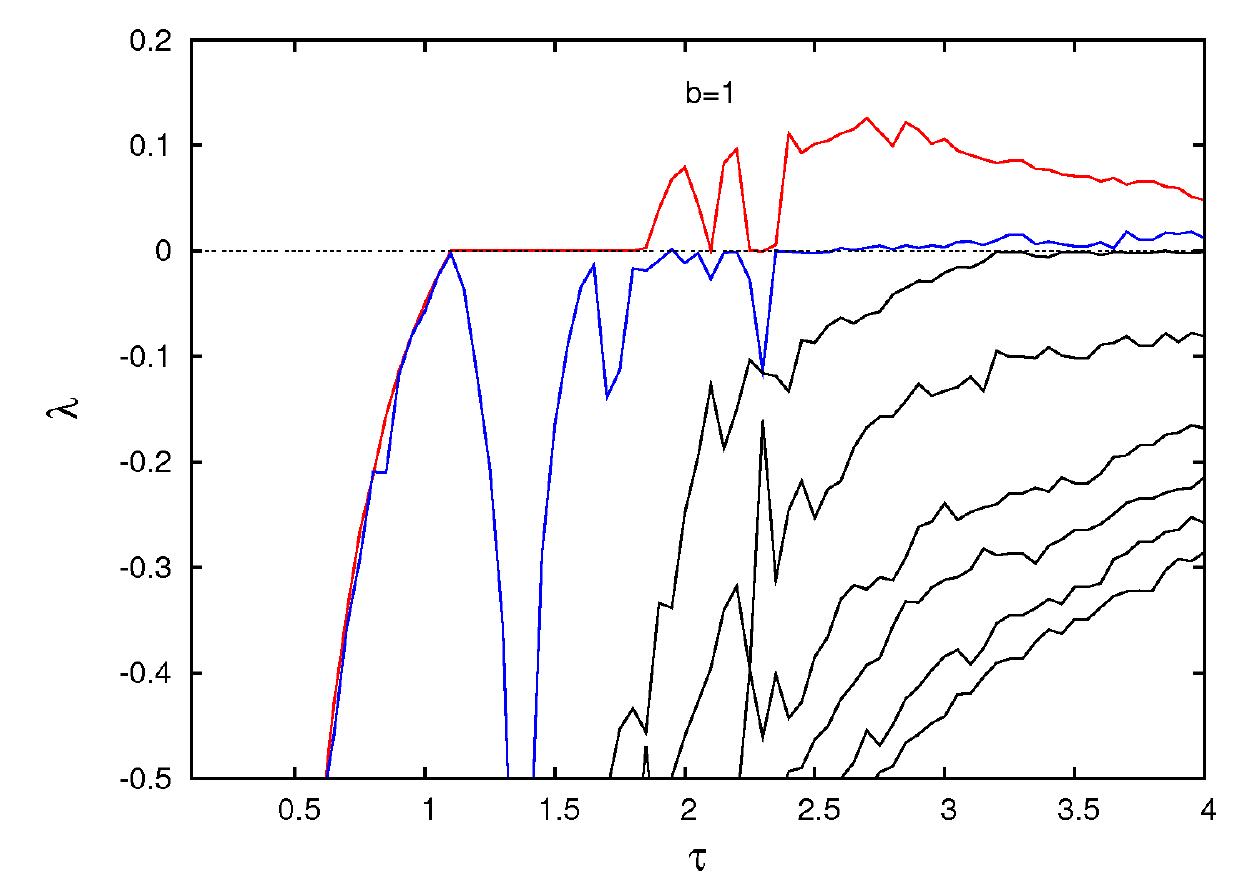}
\caption{The  first eight Lyapunov exponents ($\lambda$) with $\tau$; First two LEs become positive for $\tau \ge 3.25$ indicating hyperchaos. Other parameters are same as Fig. \ref{phtau}.}
\label{le_tau}   
\end{figure}

The detailed stability and bifurcation analysis, experimental implementation and results have been reported in \cite{banerjee12}; there we have proved the existence of chaos and hyperchaos through the presence of strange attractor along with positive Lyapunov exponent \cite{len} and higher values ($>3$) of Kaplan--York dimension. Also we discussed the distinct features of this time-delay circuit over the existing piece-wise-linear nonlinearity based time-delay circuits \cite{for}. For better understanding of the coupled case, let us discuss the important characteristics of the the system. Following parameter values have been used in \cite{banerjee12}: $a=1$, $n=2.2$, $m=1$, $l=10$. It has been shown that, for $b=1$, if one varies $\tau$, for $\tau \ge 1.102$, the fixed point loses its stability through Hopf bifurcation. At $\tau=1.65$, limit cycle of period-1 becomes unstable and a period-2 (P2) cycle appears. Further period doubling occurs at $\tau=1.79$ (P2 to P4). Through a period doubling sequence, the system enters into the chaotic regime at $\tau=1.84$. With further increase of $\tau$, at $\tau=2.60$, the system shows the emergence of hyperchaos. The system  shows a double scroll at $\tau \approx 3.24$. Phase plane  representation in the representative $x-x(t-\tau)$ plane for different $\tau$ is shown in Fig.\ref{phtau}, which shows the following characteristics: period-1 ($\tau=1.40$), period-2 ($\tau=1.72$), chaos ($\tau=1.94$), and double scroll hyperchaos ($\tau=3.6$). Figure \ref{le_tau} shows the Lyapunov exponent spectrum of the system in the $\tau$ parameter space. It is noteworthy that for a proper choice of $b$, the system shows chaos and hyperchaos even for a small time delay; e.g. for $b=1$ one has chaos for $\tau\approx1.84$ and hyperchaos for $\tau\ge2.60$. This makes the circuit implementation of the system easier and also makes it superior for the possible applications in communication system.
\section{Experiment}
\label{sec:3}
\subsection{Electronic circuit realization}
\begin{figure}
 \includegraphics[width=.49\textwidth]{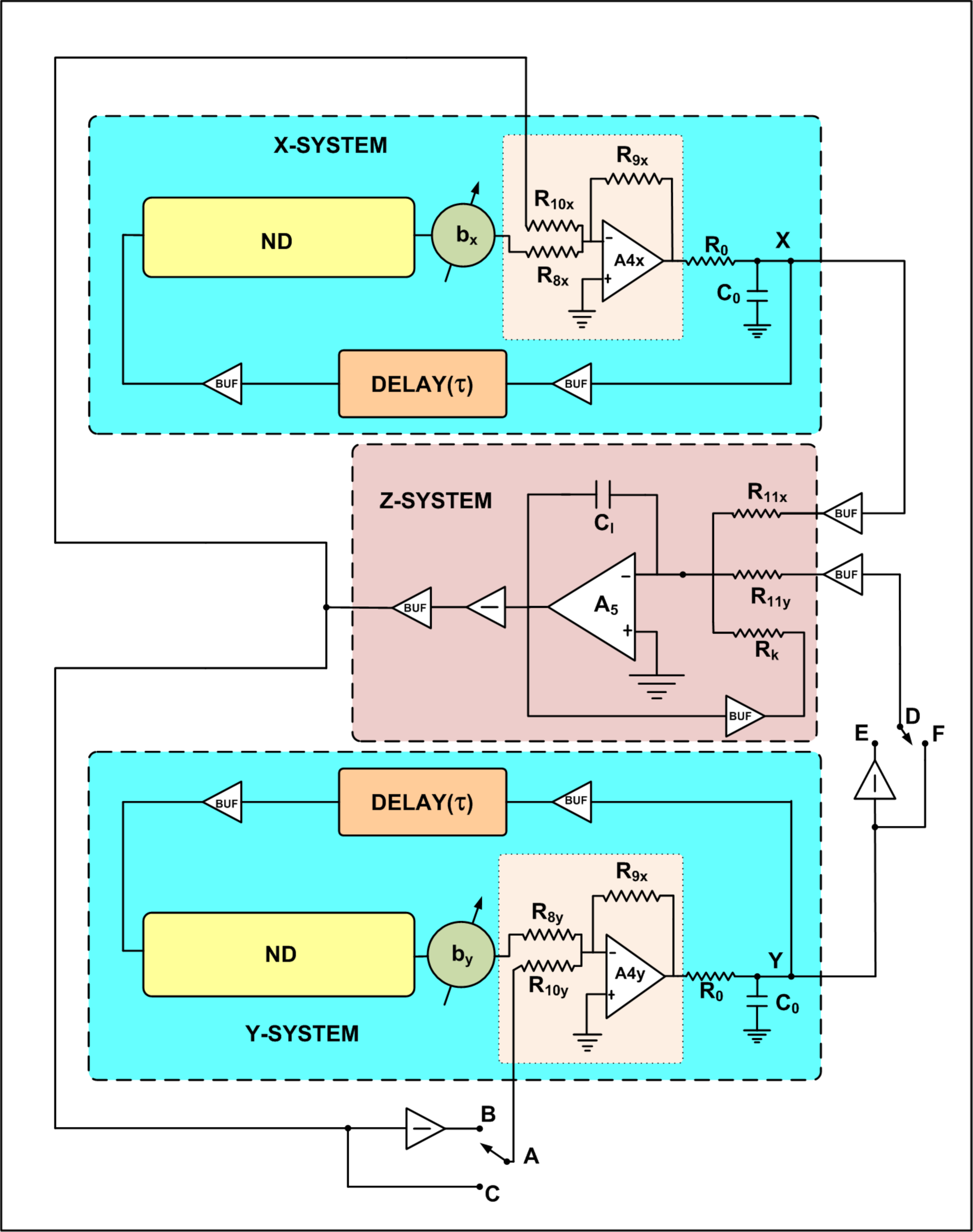}
 \caption{Experimental circuit diagram of the coupled system. $R_0=1$ k{\ohm}, $C_0=0.1$ $\mu$F, $C_I=0.1$ $\mu$F. Buffers are designed with the unity gain non-inverting operational amplifiers. $b_{x,y}$ are amplifiers and ND represents nonlinear device (see Fig. \ref{nlckt}). A4x,y op amps are used as inverting adder and A5 op amp is used as inverting integrator. All the opamps are TL 074. $R_k=1$ k{\ohm}, $R_{8x,y}=R_{9x,y}=1$ k{\ohm}. In experiment the following condition is always met: $R_{10x}=R_{10y}$ and $R_{11x}=R_{11y}$.} 
\label{ckt}
\end{figure}
\begin{figure}
 \includegraphics[width=.49\textwidth]{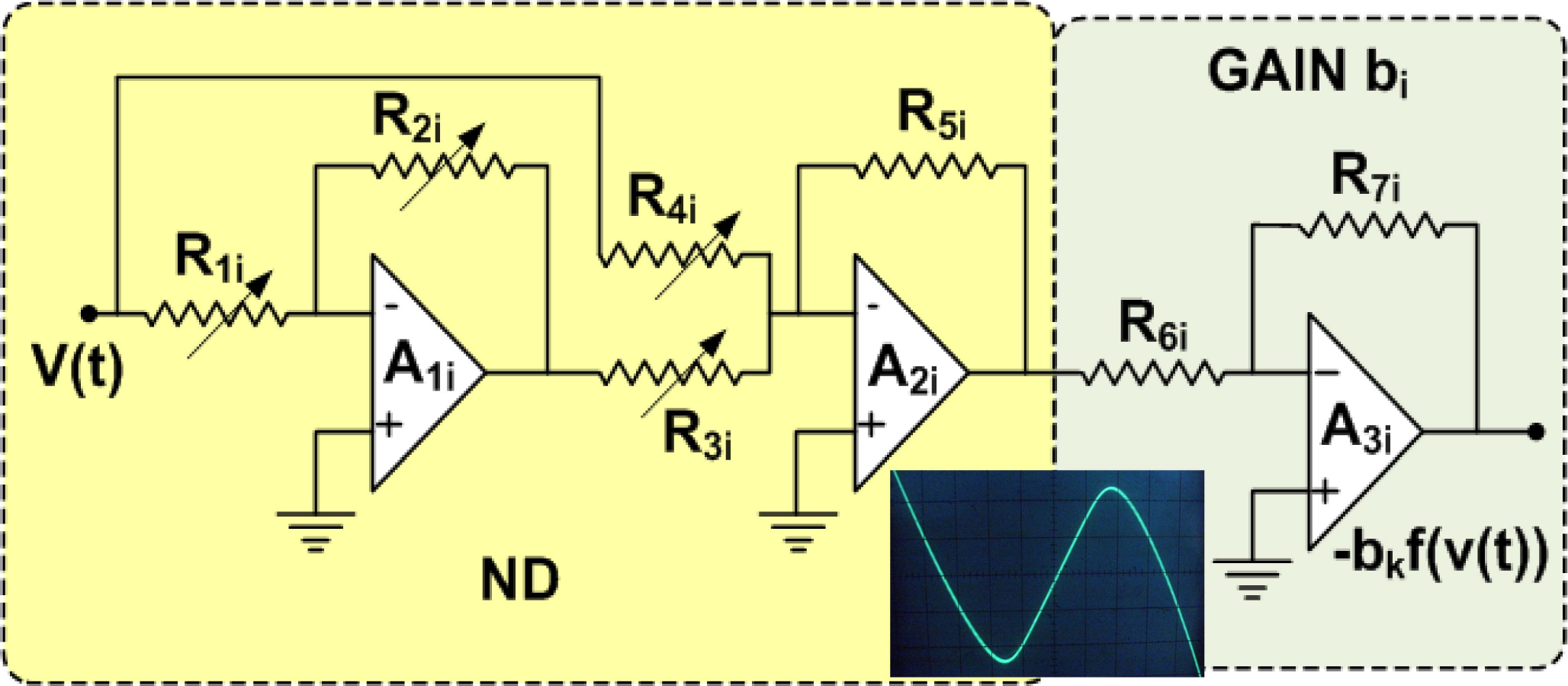}
 \caption{Nonlinear Device (ND) along with the gain block ($b_i$). A1i-A3i ($i=x,y$) are op-amps (TL 074), $R_1=1.26$ k\ohm, $R_2=19.29$ k{\ohm}, $R_3=52.81$ k{\ohm}, $R_4=6.73$ k{\ohm}, $R_5=10$ k{\ohm}, $R_6=1$ k{\ohm}. Inset shows the experimental oscilloscope trace of the nonlinearity produced by the ND.}
 \label{nlckt}
\end{figure}

We implement the coupled system given by \eqref{dr} in an analog electronic circuit. Figure \ref{ckt} shows the representative diagram of the experimental circuit. The proposed circuit consists three distinct parts, namely, the {\em $x$-system}, the {\em $y$-system} and the {\em $z$-system} or the {\em environment}. The circuit of nonlinear device (ND) of each systems is given in Fig. \ref{nlckt}; delay block is realized by using active all-pass filters (APF). To achieve the indirect coupling via environment, we feed the outputs from the $R_0-C_0$ junctions of the $x$- and $y$- systems to the inverting terminal of the op-amp A5, which acts as an integrator, through buffers (for impedance matching) and resistors $R_{11x}$ and $R_{11y}$, which determine the parameter $\epsilon_2$. The voltage from the $R_0-C_0$ terminal of the $y$-system can be inverted for the in-phase coupling (i.e. $\beta_1=1, \beta_2=-1$) by the use of an unity gain inverter  and connecting the points ``D" and ``E". Whether for inverse-phase coupling ($\beta_1=1, \beta_2=1$), terminals "D" and "F" will be connected. The output of the integrator A5 is fed into the inverting terminal of it through a buffer and a resistance $R_k$; $R_k$ will determine the parameter $\kappa$. Further, the output of the integrator is passed through an unity gain inverter and a buffer, and it is distributed in two ways: {\bf (i)} it is directly added to the $x$-system by the use of an inverting adder (A4x) through $R_{10x}$, that determines the coupling strength $\epsilon_1$, and {\bf (ii)} it is passed through an inverter (for $\beta_1=1, \beta_2=-1$) or directly connected (for $\beta_1=1, \beta_2=1$) to inverting adder A4y through resistor $R_{10y}$ that also determines $\epsilon_1$; thus we always kept $R_{10x}=R_{10y}$ and $R_{11x}=R_{11y}$. 

Let, $V_1(t)$ be the voltage drop across the capacitor $C_0$ of the low-pass filter section of the $x$-system and that of the $y$-system be $V_2(t)$. Also, let the output of the integrator A5 be $V_3(t)$, which represents the {\it environment}. Thus the following equations represent the time evolution of the circuit: 
\begin{subequations}\label{xyzsys}
\begin{align}
R_0C_0\frac{dV_1(t)}{dt}&=-V_1(t)+\frac{R_{9x}}{R_{8x}} b_1f({V_{1}}_{T_{D}})\nonumber\\
                        &~~~~~~~~~~~~~~~~~~~~+\frac{R_{9x}}{R_{10x}}\beta_1 V_3(t),\label{xsys}\\
R_0C_0\frac{dV_2(t)}{dt}&=-V_2(t)+\frac{R_{9y}}{R_{8y}} b_2f({V_{2}}_{T_{D}})\nonumber\\
                        &~~~~~~~~~~~~~~~~~~~~+\frac{R_{9y}}{R_{10y}}\beta_2 V_3(t),\label{ysys}\\
C_I\frac{dV_3(t)}{dt}&=-\frac{1}{R_k}V_3(t)-\frac{1}{R_{11x}}\beta_1V_1(t)\nonumber\\
                     &~~~~~~~~~~~~~~~~~~~~~~~~~~~-\frac{1}{R_{11y}}\beta_2V_2(t).\label{zsys}
\end{align}
\end{subequations}
Here, $b_1,b_2=\frac{R_{7i}}{R_{6i}}$ is the gain of the amplifier A3i ($i=x,y$) (Fig.\ref{nlckt}), $\beta_1=1$, and $\beta_2=\mp 1$, depending upon connection topology. $f({V_j}_{T_{D}})\equiv f(V_j(t-T_D)\equiv f({V_j}_{\tau})$, ($j=1,2$), is the nonlinear function representing the output of the Nonlinear Device (ND) of Fig.\ref{nlckt}, in terms of the input voltage ${V_j}_{\tau}$. $T_{D}$ is the time delay produced by the delay block. Also, we choose $R_{8x,y}=R_{9x,y}=1$ k{\ohm}.
\begin{figure}
 \includegraphics[width=.49\textwidth, height=0.22\textheight]{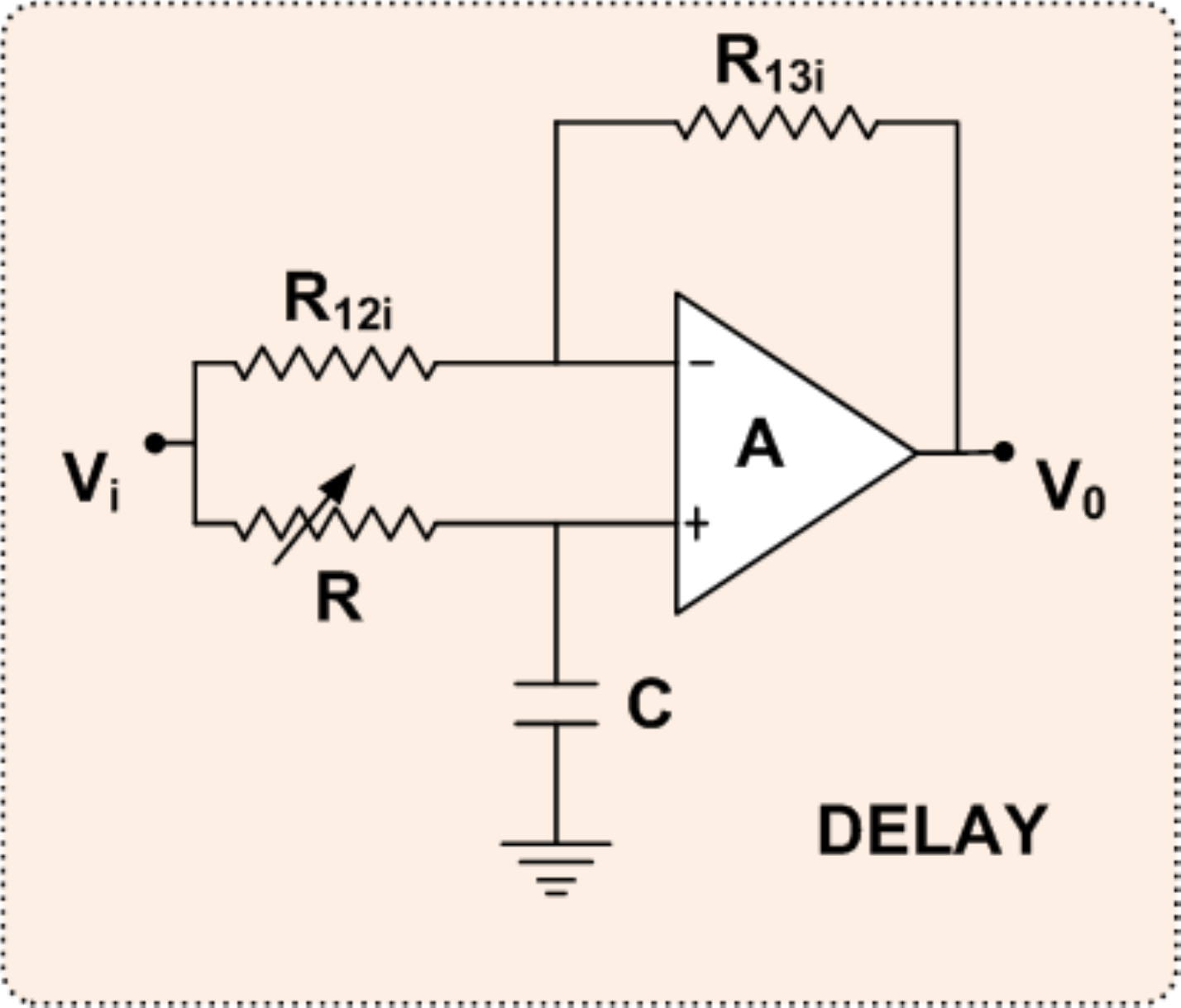}
 \caption{Active first-order all-pass filter. $R_{12}=R_{13}=2.2$ k{\ohm}, $C=10$ nF. ``A" represents TL 074 opamp.}
 \label{apf}
\end{figure}
In  \cite{banerjee12}, we have reported that the nonlinearity of nonlinear device given by Fig. \ref{nlckt} has the following form:
\begin{equation}\label{ex3}
\begin{split}
f(V_{j_{T_D}})&=-\frac{R_{5i}}{R_{4i}}V_{j_{T_D}}\\
                        &~~+\frac{R_{5i}}{R_{3i}}\beta V_{sat}\tanh\biggl(w\frac{R_{2i}}{R_{1i}}\frac{V_{j_{T_D}}}{V_{sat}}\biggr).
\end{split}
\end{equation}
Here $\beta$ and $w$ are the scaling factors that depend upon the non-ideal nature and asymmetry of the op amps. In general, for smaller input voltages, $\beta \approx1$ and $w\approx1$. $V_{sat}$ is the saturation voltage of the opamp. The variable delay element is realized by a first order {\em all-pass filter} (APF) (Fig.\ref{apf}) \cite{sed}. The APF has the following transfer function:
\begin{equation}
T(s)=-a_1\frac{s-\omega_0}{s+\omega_0},
\end{equation}
with flat gain $a_1=1$ (determined by $R_{12i}$ and $R_{13i}$), and $\omega_0=1/CR$ is the frequency at which the phase shift is $\pi/2$. Since it has an almost linear phase response, thus each APF block contributes a delay of $T_D\approx RC$. So  $n$ blocks produce a delay of $T_D=nRC$ ($n=1,2,\dots$). By simply changing the resistance $R$, one can vary the amount of delay; thus one can control the resolution of the delay line (the same technique of implementing delay line in the system and the coupling path has been used in \cite{{tangs},{tanls}}, which differs and is actually advantageous over the conventional techniques used in \cite{{lak11},{lakpsexpt},{zls}}).

Let us define the following dimensionless variables and parameters: $t=\frac{t}{R_0C_0}$, $\tau=\frac{T_D}{R_0C_0}$, $x=\frac{V_1(t)}{V_{sat}}$, $x_{\tau}=\frac{V_{1_{T_D}}}{V_{sat}}$, $y=\frac{V_2(t)}{V_{sat}}$, $y_{\tau}=\frac{V_{2_{T_D}}}{V_{sat}}$, $z=\frac{V_3(t)}{V_{sat}}$, $\frac{R_{5i}}{R_{4i}}=n_1$, $\beta \frac{R_{5i}}{R_{3i}}=m_1$, $w\frac{R_{2i}}{R_{1i}}=l_1$, $b_{1,2}=\frac{R_{7i}}{R_{6i}}$, $\epsilon_1=\frac{R_{9i}}{R_{10i}}$, $\kappa=\frac{R_0C_0}{R_kC_I}$, $\frac{\epsilon_2}{2}=\frac{R_0C_0}{R_{11i}C_I}$, where $i=x,y$. To make the time-scale of the $x,y$-systems and the $z$-system equal we use $C_I=C_0$. Now, we get $\kappa=\frac{R_0}{R_k}$, $\frac{\epsilon_2}{2}=\frac{R_0}{R_{11i}}$. Thus, to make $\epsilon_1=\epsilon_2$, we have to use $R_{11i}=2R_{10i}$ (since $R_0=R_{9i}=1$ k{\ohm}).

Now, the equations \eqref{xyzsys} and \eqref{ex3} can be reduced to the following dimensionless, coupled, first-order, retarded type nonlinear delay differential equations:
\begin{subequations}\label{xyzdl}
\begin{align}
\frac{dx}{dt}&=-x(t)+b_1f(x_{\tau})+\epsilon_1\beta_1 z,\label{xdl}\\
\frac{dy}{dt}&=-y(t)+b_2f(y_{\tau})+\epsilon_1\beta_2 z,\label{ydl}\\
\frac{dz}{dt}&=-\kappa z-\frac{\epsilon_2}{2}\big(\beta_1 x+\beta_2 y\big).\label{zdl}
\end{align}
\end{subequations}
where,
\begin{equation}\label{ex6}
f(u_{\tau}) =  -n_{1}u_{\tau}+m_1\tanh(l_1u_{\tau}),
\end{equation}
where $u\equiv x,y$.\\
It is worth noting that equations \eqref{xyzdl} (along with \eqref{ex6}) are  equivalent to equations \eqref{dr} (along with \eqref{hw1}) with $a=1$, and appropriate choices of $n_1$, $m_1$ and $l_1$.
\subsection{Experimental results}
The coupled system is designed in hardware level on a bread board using IC TL074 opamps (JFET quad opamps) with $\pm15$ volt power supply. Capacitors and resistors have $5\%$ tolerance . The resistance values used in the circuit for both the $x$- and $y$-systems are: $R_1=1.26$ k{\ohm}, $R_2=19.29$ k{\ohm}, $R_3=52.81$ k{\ohm}, $R_4=6.73$ k{\ohm}, $R_5=10$ k{\ohm}, $R_6=1$ k{\ohm}. For the low pass section we used $R_0=1$ k{\ohm}  and  $C_0=0.1$ $\mu$F. Also for the $z$-system we choose $R_k=1$ k{\ohm} and $C_I=0.1$ $\mu$F.  The  nonlinearity produced by the nonlinear device of both the $x,y-$systems are kept similar in nature and is shown in Fig. \ref{nlckt} (inset). The identical active all-pass filter stages of delay line (Fig. \ref{apf}) have $R_{12}=R_{13}=2.2$ k{\ohm}, $C=10$ nF and a variable resistance $R$. Here our main concern is to study the synchronization phenomena by varying the coupling strengths  keeping the other system design-parameters same for the two systems. 
\begin{figure}
\includegraphics[width=.49\textwidth]{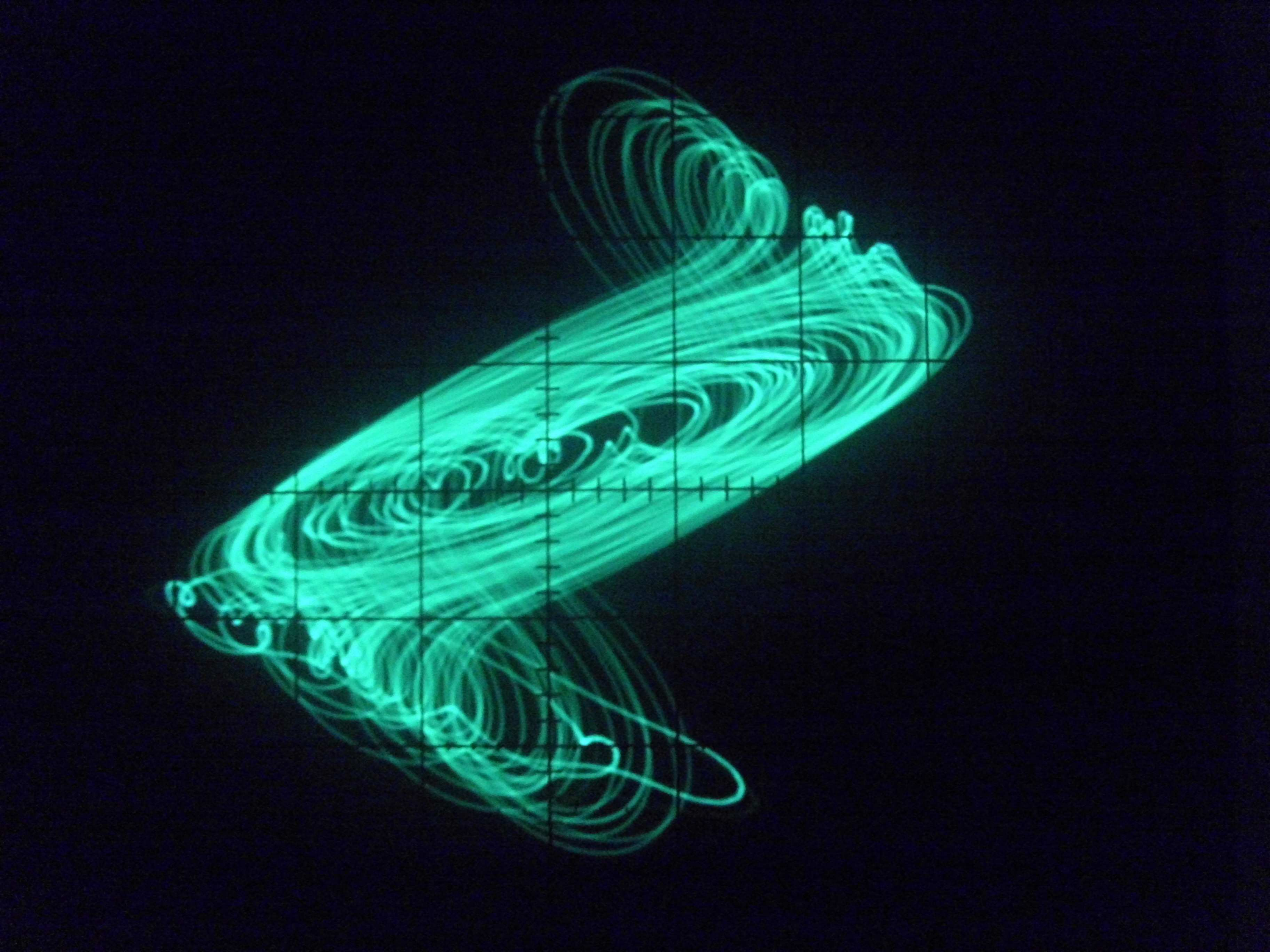}
\includegraphics[width=.49\textwidth]{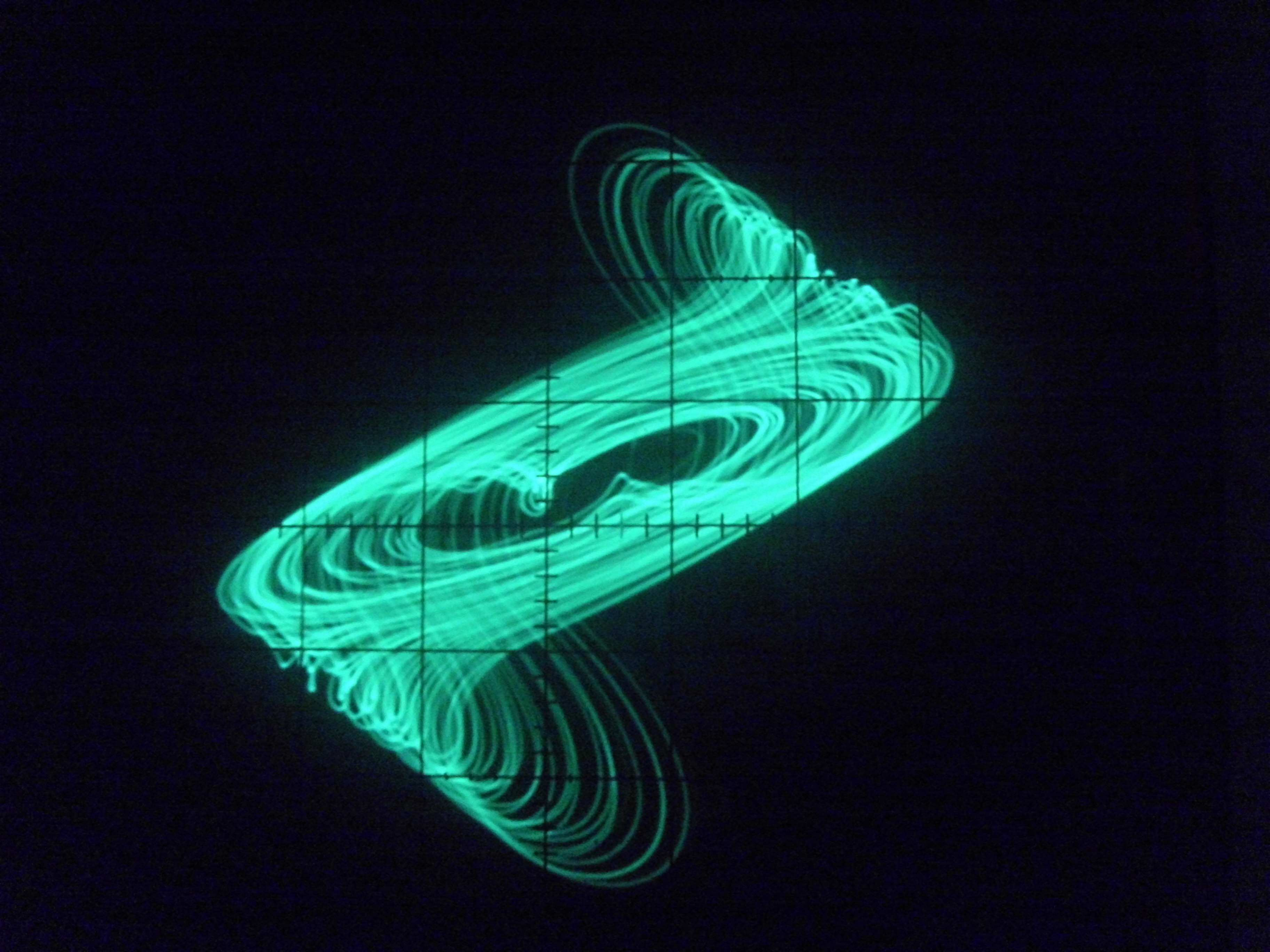}
\caption{The individual attractors of the uncoupled systems. Upper panel: $x$-system,  $V_1(t)-V_1(t-T_D)$ space, lower panel: $y$-system, $V_2(t)-V_2(t-T_D)$ space in  the hyperchaotic regime. $R_{7x,y}=2.1$ k{\ohm} (for other parameter values see text). Oscilloscope scale divisions: $x$ -axis: $0.5$ v/div, $y$-axis: $0.5$ v/div.}
\label{xyexpt}
\end{figure} 
\begin{figure}
 \includegraphics[width=.49\textwidth]{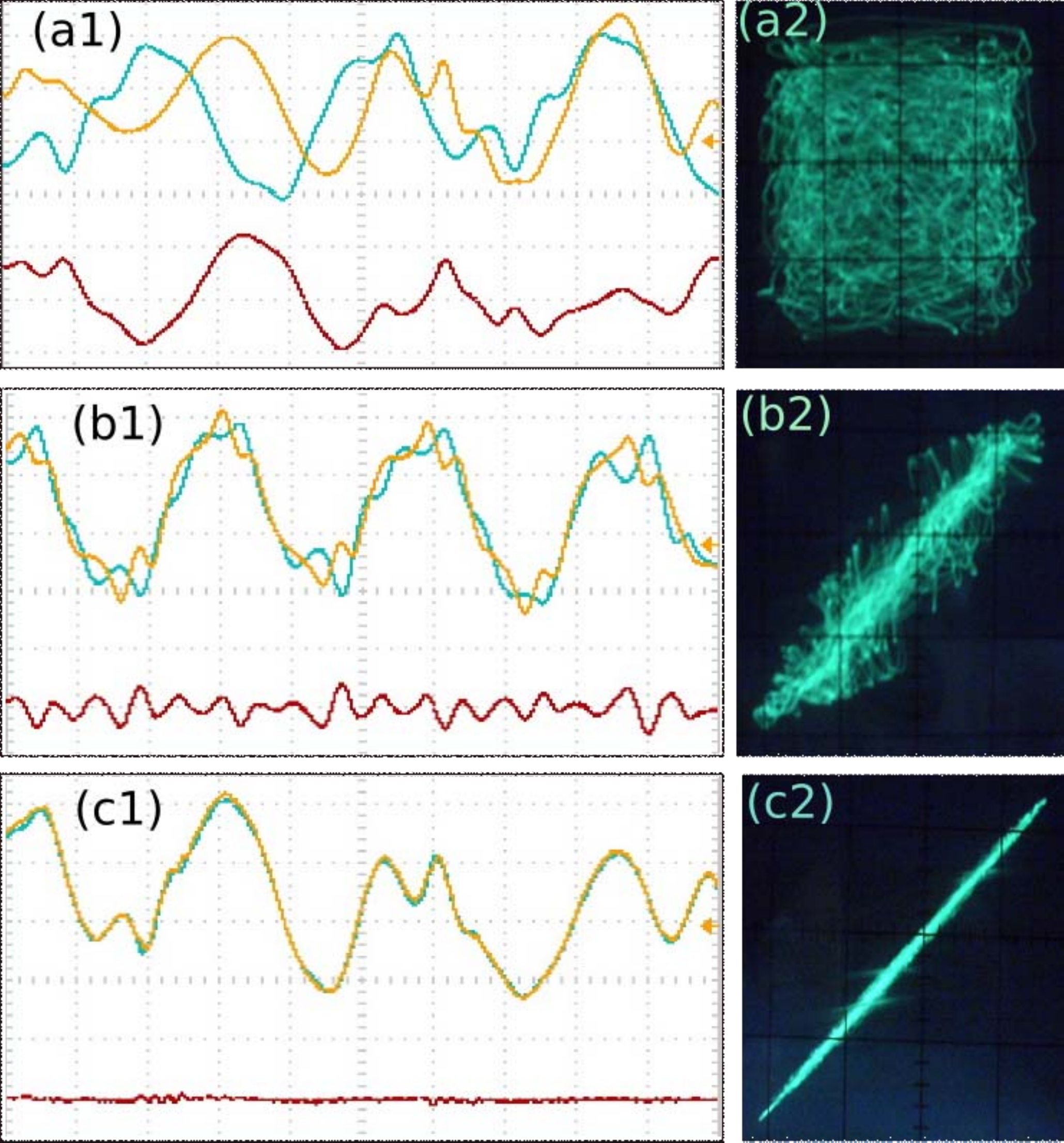}
 \caption{$\beta_1=1$ and $\beta_2=-1$: Experimental time series plot of the $x$-system $V_1(t)$ (yellow) and the $y$-system $V_2(t)$ (blue) in the {\em hyperchaotic} Regime, lower trace in red represents the error signal $(V_1(t)-V_2(t))$: (a1) unsynchronized state, (b1) in-phase synchronization (c1) complete synchronization. The corresponding phase plane plots are shown in (a2), (b2), and (c2), respectively. (For parameter values see text). For (a1), (b1), and (c1): $x$-axis: 25$\mu$s/div, $y$-axis: $1.25$ v/div. For (a2), (b2), and (c2): $x$-axis: $1$ v/div, $y$-axis: $1$ v/div. (Color figure online).}
\label{expt_d}
\end{figure}
\begin{figure}
 \includegraphics[width=.49\textwidth]{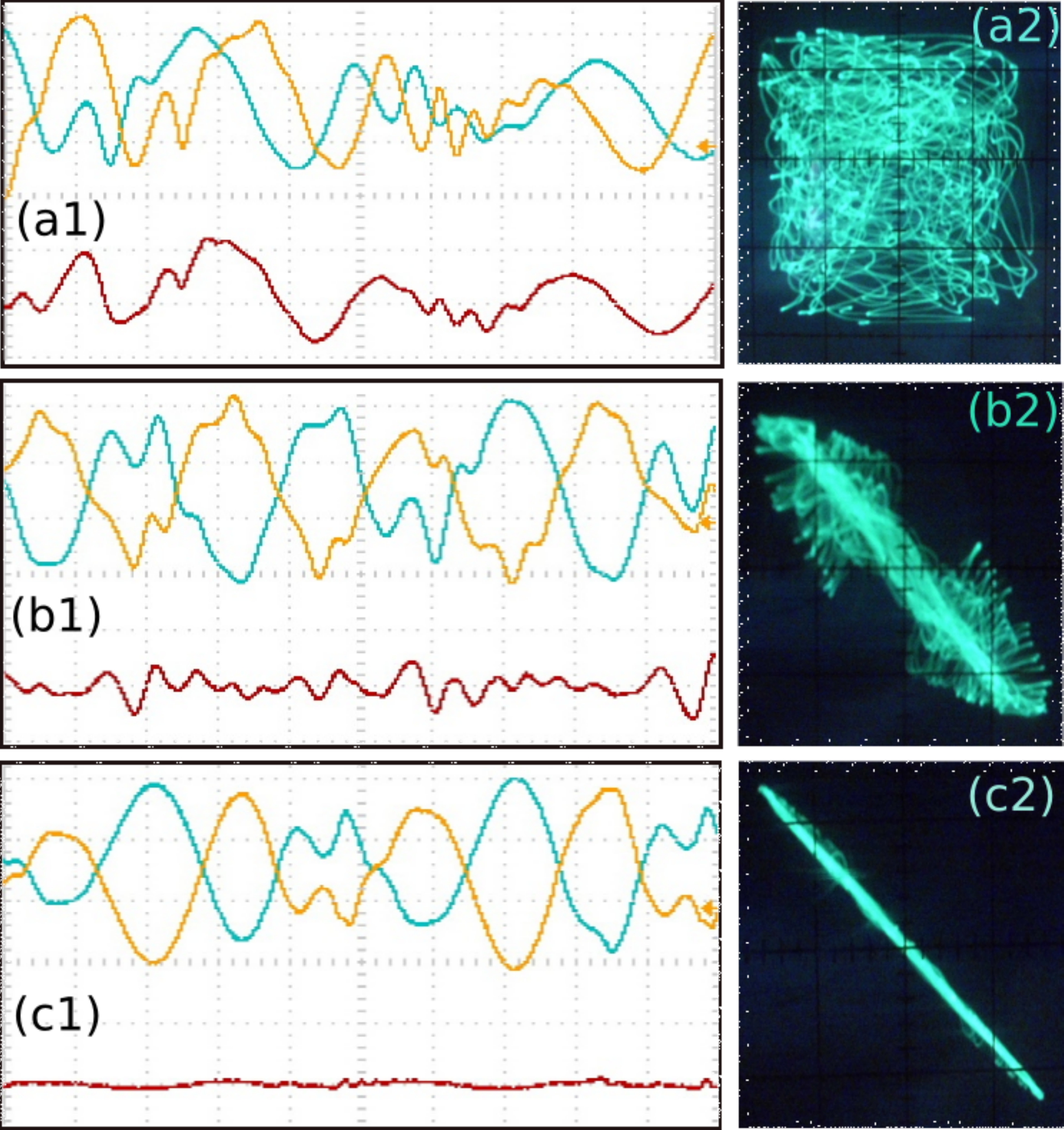}
 \caption{$\beta_1=1$ and $\beta_2=1$: Experimental time series plot of the $x$-system $V_1(t)$ (yellow) and the $y$-system $V_2(t)$ (blue) in the {\em hyperchaotic} Regime, lower trace in red represents the error signal $(V_1(t)+V_2(t))$: (a1) unsynchronized state, (b1) inverse-phase synchronization (c1) anti-synchronization. The corresponding phase plane plots are shown in (a2), (b2), and (c2), respectively. (For parameter values see text). For (a1), (b1), and (c1): $x$-axis: 25 $\mu$s/div, $y$-axis: $1.25$ v/div. For (a2), (b2), and (c2): $x$-axis: $1$ v/div, $y$-axis: $1$ v/div. (Color figure online).}
\label{expt_in}
\end{figure}   
\subsubsection{Real time waveform and phase plane plots}
To observe the real time waveform and phase plane plots we fix the feedback delay to the value $\tau=3.6$ (by choosing four stages of APF delay blocks with first three stages having $R=10$ k{\ohm} and the last one has $R=6$ k{\ohm}). We choose $R_7\approx 2.1$ k{\ohm} to keep both the systems in the hyperchaotic regime. Figure \ref{xyexpt} shows the hyperchaotic attractor of the $x-$ system (upper panel) and $y-$systems (lower panel) in the uncoupled conditions. In the following cases, we consider the coupling resistors for which $\epsilon_1=\epsilon_2$.\\ 
{\bf (i)} For $\beta_1=1$ and $\beta_2=-1$: This condition is satisfied by connecting the points ``A" to ``B", and ``D" to ``E", in the Fig. \ref{ckt}. Figure \ref{expt_d} shows three distinct cases: {\bf(a)} The $x$- and $y$-systems evaluate independently in time for large values of coupling resistance. For example, when we keep $R_{10x}=R_{10y}\approx9.96$ k{\ohm}, $R_{11x}=R_{11y}=20$ k{\ohm}, both the systems are unsynchronized. Fig. \ref{expt_d}(a1) shows the time series of the $x$-system (yellow) and the $y$-system (blue), and the red trace in the lower portion of this plot shows the difference ($V_1(t)-V_2(t)$) (call it the ``error") of the two systems captured by a digital storage oscilloscope (DSO) (Tektronix TDS2002B, $60$ MHz, 1 GS/s). It can be noticed that the amplitude of the error signal is of the same order as the original signals $V_1(t)$ and $V_2(t)$) indicating unsynchronized states. Fig. \ref{expt_d}(a2) shows the corresponding phase-plane plot in $V_1(t)-V_2(t)$ space that confirms the asynchronous condition. {\bf (b)} If we decrease the coupling resistors (i.e., increase the coupling strength $\epsilon$), the coupled systems show in-phase synchronization.  This is shown for $R_{10x}=R_{10y}\approx1.57$ k{\ohm}, $R_{11x}=R_{11y}\approx3.2$ k{\ohm} in  Fig.~\ref{expt_d}(b1)(b2); the time series in Fig.~\ref{expt_d}(b1) shows that the waveforms are in phase but their amplitude levels are still uncorrelated in time scale. Also the amplitude of error signal is much reduced here. Fig.~\ref{expt_d}(b2) shows the corresponding phase-plane diagram where we can see that the system dynamics now wanders around the $45^\circ$ diagonal line . These two indicate the in-phase synchronization of the $x$- and the $y$-systems. {\bf (c)} Further reduction in the coupling resistors results in complete synchronization. Fig. \ref{expt_d} (c1)(c2) shows this for $R_{10x}=R_{10y}\approx0.56$ k{\ohm} and $R_{11x}=R_{11y}\approx1.2$ k{\ohm}.  Fig. \ref{expt_d}(c1) and (c2) shows the real time and phase-plane plots of the systems, respectively. The error line in lower portion of Fig. \ref{expt_d}(c1) indicates that the two waveforms are equal in phase and amplitude. The phase-plane plot in Fig. \ref{expt_d}(c2) showing $45^\circ$ inclination with both the axes confirms complete synchronization. 

{\bf(ii)} For $\beta_1=\beta_2=1$: This condition is achieved when one connects the points ``A" to ``C" and ``D" to ``F", in Fig. \ref{ckt}. In this case also we get three distinct situations: {\bf (a)} The $x$- and $y$- systems evaluate independently in time for larger coupling resistance. Figure \ref{expt_in}(a1)(a2) show this for $R_{10x}=R_{10y}\approx9.93$ k{\ohm}, $R_{11x}=R_{11y}=20$ k{\ohm}. Fig. \ref{expt_in}(a1) shows the time-series of the $x$- and $y$-systems, and that in the lower portion of the plot shows the sum of the two waveforms $(V_1(t)+V_2(t))$. Fig. \ref{expt_in}(a2) shows the corresponding phase-plane plot. The sum and the phase-plane plot confirm that the systems are not synchronized. {\bf (b)} Lowering of coupling resistances results in inverse-phase synchronization. This is shown in Fig. \ref{expt_in}(b1) (b2)  for $R_{10x}=R_{10y}=2$ k{\ohm}, $R_{11x}=R_{11y}\approx4.1$ k{\ohm}. Fig. \ref{expt_in}(b1) shows the time-series and the sum $(V_1(t)+V_2(t))$, and Fig. \ref{expt_in}(b2) shows the corresponding phase-plane plot. One can see that $V_1(t)$ and $V_2(t)$ are in the phase-inverted mode but their amplitudes do not correlate. {\bf (c)} At further lower coupling resistance values the systems show anti-synchronization; Fig. \ref{expt_in}(c1) and (c2) shows the real time and phase-plane plots, respectively, for $R_{10x}=R_{10y}\approx0.3$ k{\ohm}, $R_{11x}=R_{11y}\approx0.7$ k{\ohm}. Here one can see that the two systems have $\pi$ phase shift and also their amplitude levels are same. The phase-plane plot shows that the system dynamics lives in a diagonal line making an angle $135^\circ$ with both the axes. This confirms the occurrence of anti-synchronization. 
\subsubsection{Generalized autocorrelation function and correlation of probability of recurrence}
\label{sub:cpr}
It is seen from numerical simulations (Fig.~\ref{phtau}) and experimental results (Fig.~\ref{xyexpt}) that the attractor of the system under study is not phase-coherent. As there exists no general technique of finding phase of a phase-incoherent hyperchaotic attractor, we use two dynamical measures of phase synchronization (PS) proposed in Ref. \cite{{cpr1},{cpr2}}, namely, generalized autocorrelation function ($P(t)$) and  correlation of probability of recurrence (CPR). These two measures have effectively been used in the context of chaotic phase synchronization of coupled time-delayed systems (e.g. \cite{lakpsexpt}, \cite{zls}, \cite{kurth_ikeda}, etc.). Here, we compute $P(t)$ and CPR of the coupled system {\it experimentally} using the time-series data acquired from the experimental circuit.  

The generalized autocorrelation function ($P(t)$) is defined as \cite{{cpr1},{cpr2}}
\begin{equation}\label{cpr}
P(t)=\frac{1}{N-t}\sum_{i}^{N-t}\Theta \left(\epsilon_t-\lVert X_i-X_{i+t}\rVert\right),   
\end{equation}
where $\Theta$ is the Heaviside function, $X_i$ is the $i-th$ data point in the $X$ variable, $N$ is the total number of data points, $\epsilon_t$ is a preassigned threshold value, and $\lVert . \rVert$ represents the Euclidean norm. Let, $P_1(t)$ represents the generalized autocorrelation function of the $x$-system and $P_2(t)$ be that of the $y$-system. We compute $P_1(t)$ and $P_2(t)$ from the experimental time-series data ($N=2400$) acquired using DSO (Tektronix TDS2002B, $60$ MHz, 1 GS/s). For both the cases we choose the threshold value $\epsilon_t=0.01$. 

\begin{figure}
\includegraphics[width=.49\textwidth]{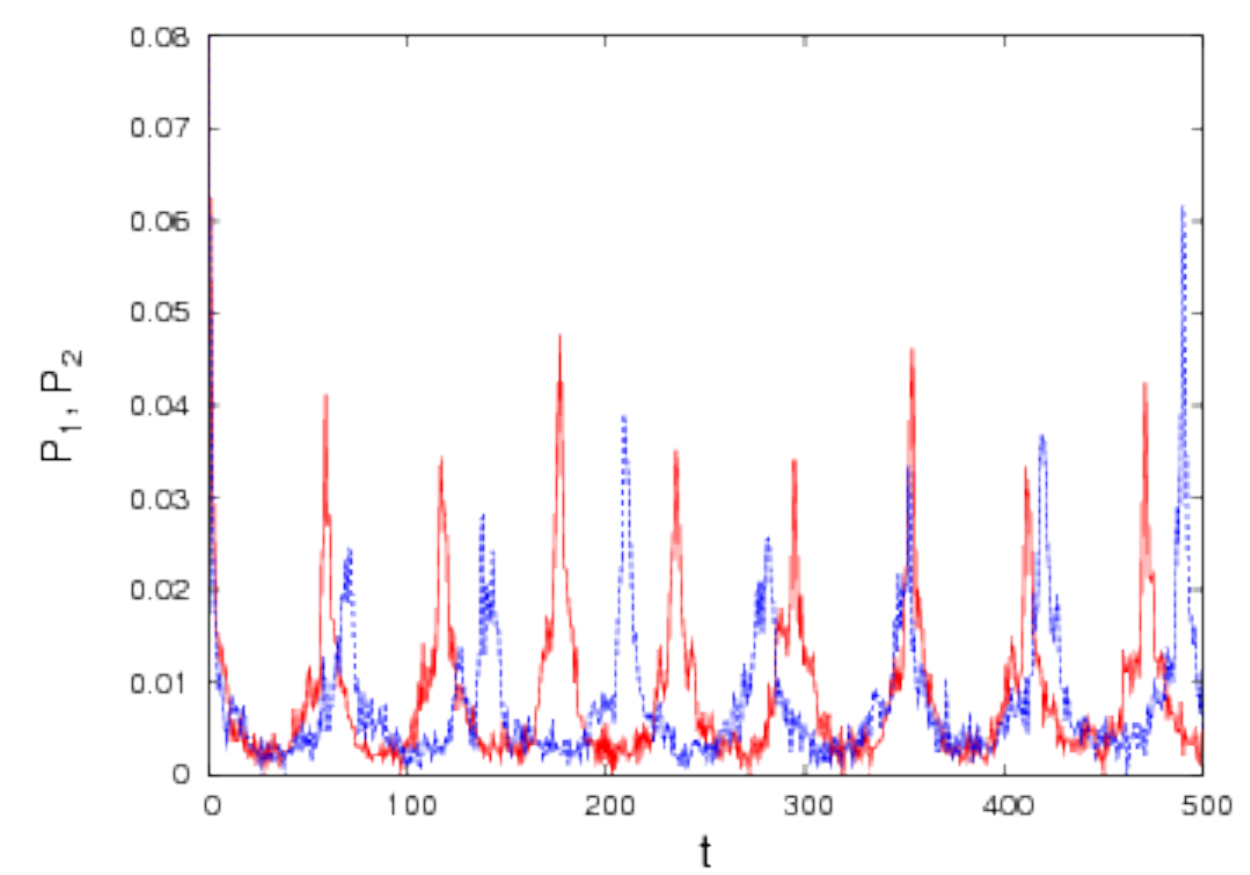}
\includegraphics[width=.49\textwidth]{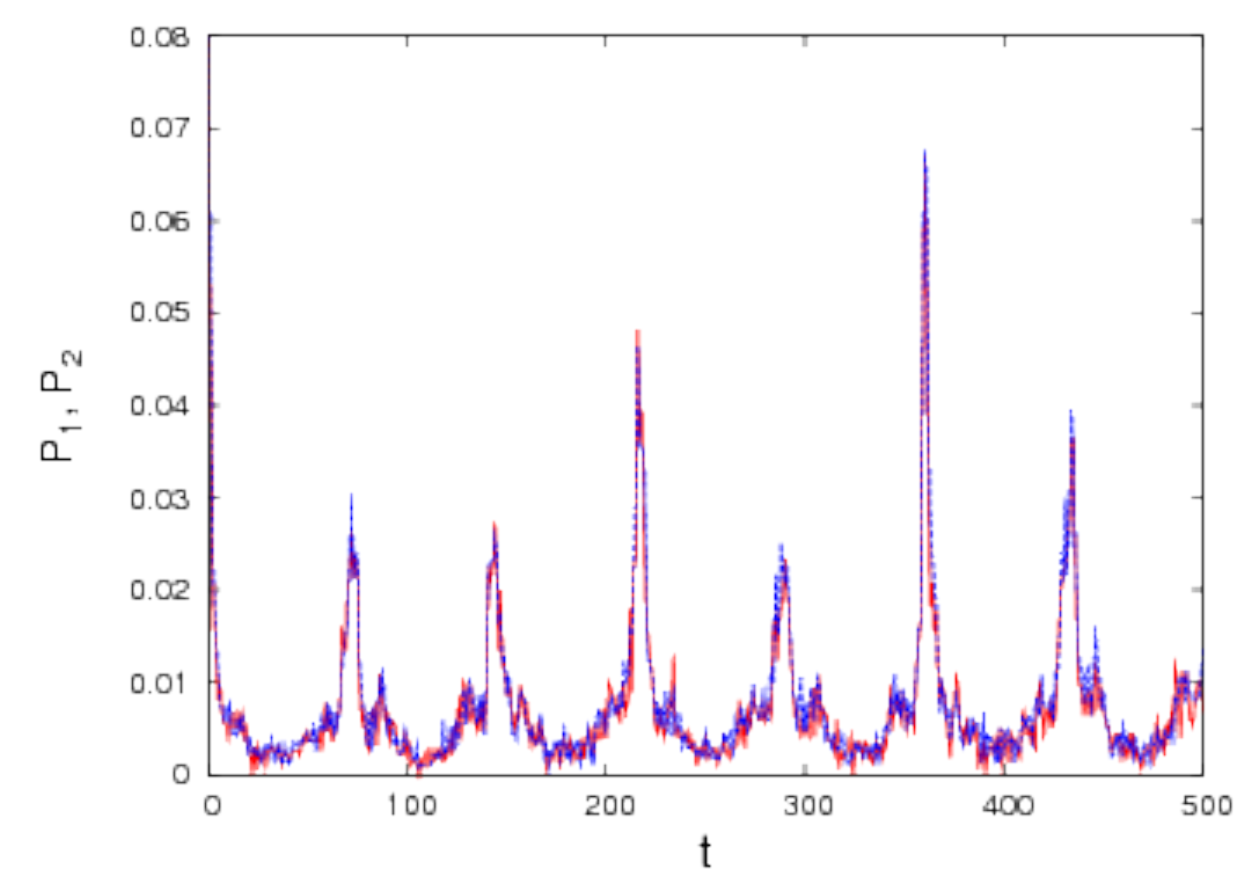}
\caption{$\beta_1=1, \beta_2=-1$: Generalized autocorrelation function ($P(t)$) computed from the experimental time-series of $x$-system ($P_1$, in red ) and $y$-system ($P_2$, in blue) in the unsynchronized state (upper panel) and in-phase synchronized state (lower panel). Notice the matching (difference) of the peaks of $P_1$ and $P_2$ in the in-phase synchronized (unsynchronized) case (for parameter values see text). (Color figure online).}
\label{dcpr_expt}
\end{figure}

\begin{figure}
\includegraphics[width=.49\textwidth]{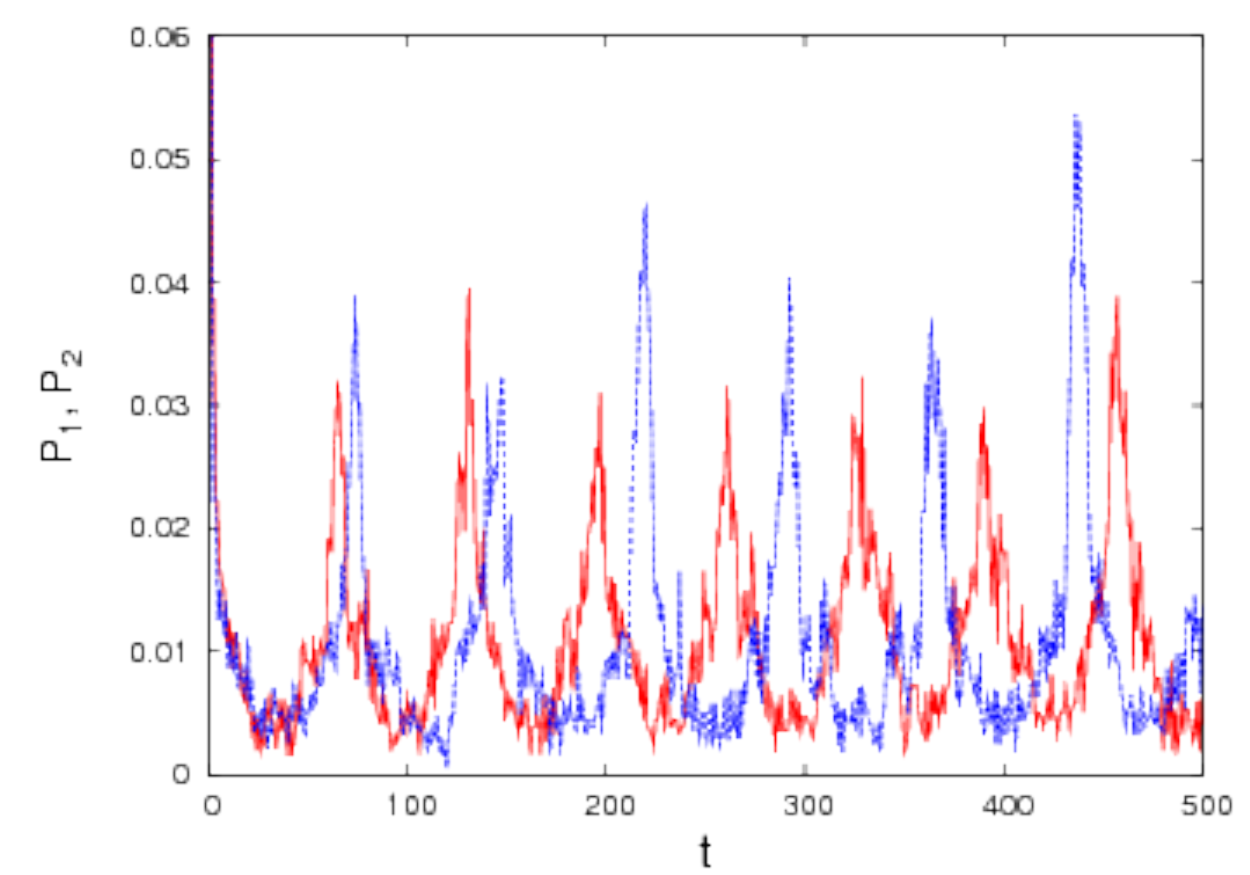}
\includegraphics[width=.49\textwidth]{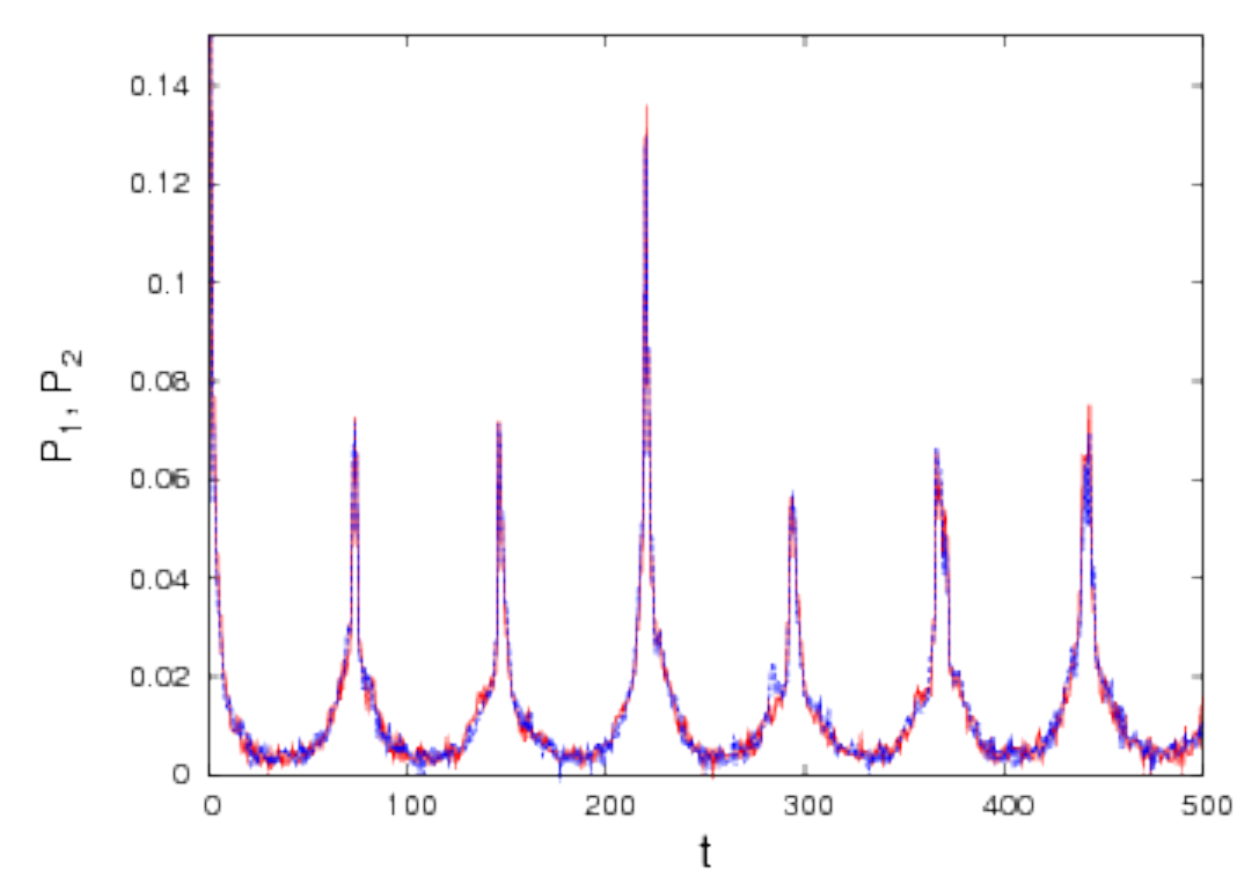}
\caption{$\beta_1=1, \beta_2=1$: Generalized autocorrelation function ($P(t)$) computed from the experimental time-series of $x$-system ($P_1$, in red) and $y$-system ($P_2$, in blue) in the unsynchronized state (upper panel) and inverse-phase synchronized state (lower panel). Notice the matching (difference) of the peaks of $P_1$ and $P_2$ in the inverse-phase synchronized (unsynchronized) case (for parameter values see text). (Color figure online).}
\label{incpr_expt}
\end{figure}

Fig.~\ref{dcpr_expt} shows $P(t)$s for $\beta_1=1, \beta_2=-1$ (i.e. in-phase coupling case), and Fig.~\ref{incpr_expt} shows the same for $\beta_1=1, \beta_2=1$ (i.e. inverse-phase coupling case). Upper-panel of  Fig.~\ref{dcpr_expt} represents the unsynchronized state for $\beta_1=1, \beta_2=-1$ at $R_{10x,y}\approx9.96$ k{\ohm}, $R_{11x,y}=20$ k{\ohm} (same parameter values as of Fig.\ref{expt_d}(a1)); similarly  Fig.~\ref{incpr_expt} (upper-panel) shows the same for $\beta_1=1, \beta_2=1$ at $R_{10x,y}\approx9.93$ k{\ohm}, $R_{11x,y}=20$ k{\ohm} (same parameter values as of Fig.\ref{expt_in}(a1)). From both the figures we can see that peaks of $P_1(t)$  does not match with that of $P_2(t)$ in the $t$-axis, confirming the fact that the phases of the two oscillators are not synchronized. Lower panels of Fig.~\ref{dcpr_expt} ($R_{10x,y}\approx1.57$ k{\ohm}, $R_{11x,y}\approx3.2$ k{\ohm}) and Fig.~\ref{incpr_expt} ($R_{10x,y}=2$ k{\ohm}, $R_{11x,y}\approx4.1$ k{\ohm}) show that the dominant peaks of $P_1(t)$ and $P_2(t)$ matches exactly in the $t$-axis confirming the phase synchronization of the coupled oscillators. Also, almost equal amplitude of the peaks in the phase synchronized cases shows the good quality of PS in the coupled systems. 

A quantitative measure of $P(t)$ is  the correlation of probability of recurrence (CPR) that was defined in Ref.~\cite{{cpr1},{cpr2}} as
\begin{equation}\label {cpr1}
CPR=\frac{\langle \bar{P_1}(t)\bar{P_2}(t)\rangle}{\sigma_1\sigma_2},
\end{equation}
${\bar{P}}_{1,2}$ present that the mean value has been subtracted, and $\sigma_{1,2}$ are the standard deviations of the $P_1(t)$ and $P_2(t)$, respectively. In the phase synchronized case, generally, CPR$\approx1$, and for the unsynchronized cases its value is appreciably smaller than 1. Using the definition of \eqref{cpr1}, we compute CPR from experimental time-series data both for the in-phase and inverse-phase cases. For in-phase synchronized case we have CPR=0.997 (parameters are same as of lower panel of Fig.~\ref{dcpr_expt}), and for  inverse-phase synchronized case we have CPR=0.985 (parameters are same as of lower panel of Fig.~\ref{incpr_expt}), which confirm the occurrence of phase synchronization in the coupled system.
\subsubsection{Concept of localized sets}
\label{sub:cls}
We use another dynamical measure for qualitative confirmation of phase synchronization called the {\it concept of localized sets}  (CLS) proposed in \cite{kurth_local}. It was shown in \cite{kurth_local} that the CLS technique is extremely useful to detect phase synchronization even when no proper measure of phase is possible. The idea of CLS in coupled oscillators is based on the fact that if one identifies a certain event in the first oscillator and then track the second oscillator at that particular event, a set $D$ will be obtained for the second oscillator; if that set spreads over the whole attractor space of the second oscillator then one may say that there is no phase correlation between the two coupled oscillators. On the contrary, if the set $D$ becomes localized to a certain zone of the attractor space then one can say that the coupled oscillators are phase synchronized.

\begin{figure}
\includegraphics[width=.49\textwidth]{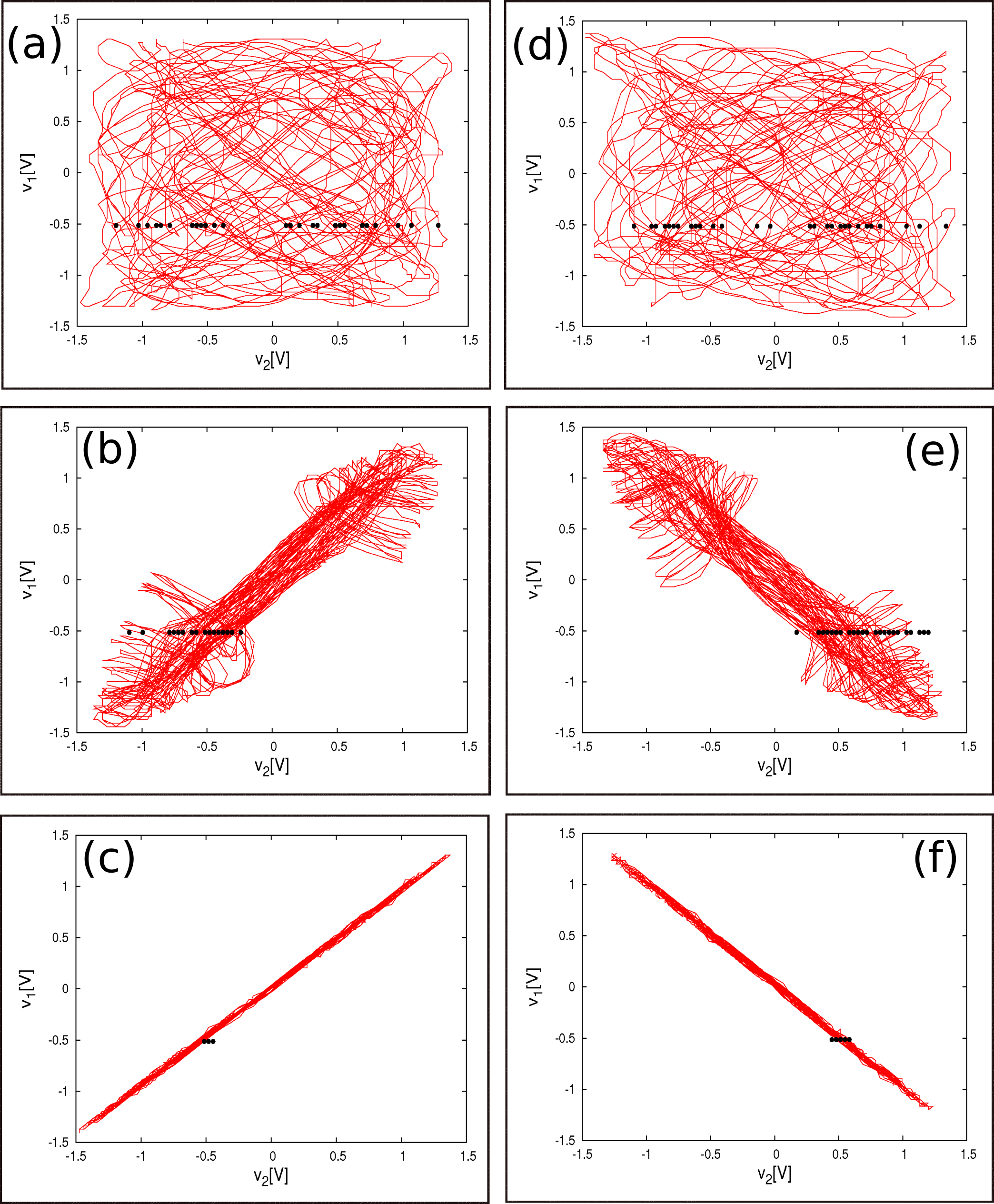}
\caption{Concept of localized sets computed from the experimental time-series: Black dots represent the set $D$ for the particular event $V_1(t)=-0.5$, the points are plotted along with $V_1(t)$ and $V_2(t)$ in the $V_1(t)-V_2(t)$ space. $\beta_1=1, \beta_2=-1$: (a) unsynchronized (b) in-phase synchronized (c) complete synchronized states. $\beta_1=1, \beta_2=1$: (d) unsynchronized (e) inverse-phase synchronized (f) anti-synchronized states. Note the localization of the set $D$ in the narrow region in $V_2(t)$ axis for the phase synchronized (b,e) and complete (anti-) synchronized cases (c,f). For parameter values see text. (Color figure online).}
\label{lset_expt}
\end{figure}

From the experimental circuit we simultaneously acquire the time-series data of $x$- and $y$-systems (with 2400 data points for each). Next, we define the event $V_1(t)=-0.5$  for the $x-$system and track the values of $V_2(t)$ from the time-series whenever that event is met. The obtained set of data values of $V_2(t)$, and the corresponding $V_1$ and $V_2$ are plotted in $V_1-V_2$ space. Figure.~\ref{lset_expt} shows the localized sets for the in-phase and inverse-phase cases at unsynchronized, phase synchronized and complete (anti-) synchronized cases. It can be seen that for the unsynchronized cases ( Fig~.\ref{lset_expt} (a) for $\beta_1=1, \beta_2=-1$  and Fig.~\ref{lset_expt} (d) $\beta_1=1, \beta_2=1$), the set $D$ (represented by black points) spreads over the whole attractor space of $V_2(t)$ (resistor values are same as used in Fig.~\ref{expt_d} (a1) ($\beta_1=1, \beta_2=-1$) and Fig.~\ref{expt_in} (a1) ($\beta_1=1, \beta_2=1$), respectively). With further decrease in $R_{10x,y}$ and $R_{11x,y}$ (that is increase in coupling strength) the set $D$ becomes localized in $V_2(t)$-space indicating the occurrence of phase synchronization; Fig.~\ref{lset_expt} (b) and (e) show this case for in-phase and inverse-phase cases, respectively (resistor values are same as used in Fig.~\ref{expt_d} (b1) ($\beta_1=1, \beta_2=-1$) and Fig.~\ref{expt_in} (b1) ($\beta_1=1, \beta_2=1$), respectively). Here, we can see that the black points representing the set $D$ are localized to a small zone of $V_2(t)$. Another interesting observation can be made from Fig.~\ref{lset_expt}, that is, for in-phase synchronized state (Fig.~\ref{lset_expt} (b)), the localized set $D$ is situated around $V_2(t)=-0.5$, which is equal to  $V_1(t)=-0.5$, whereas for inverse-phase synchronized state (Fig.~\ref{lset_expt} (e)), the localized set $D$ is situated around $V_2(t)=0.5$, which is opposite of $V_1(t)=-0.5$ (indicating a $\pi$ phase shift between $V_1(t)$ and $V_2(t)$). Finally, with further decrease in coupling resistance (i.e. increase in coupling strength) we observe that the set $D$ becomes localized to a very narrow range of $V_2(t)$ that indicates complete synchronization (Fig.~\ref{lset_expt} (c)) and anti-synchronization (Fig.~\ref{lset_expt} (f)).
\section{Linear stability analysis}
\label{sec:4}
In this section we explore the linear stability of synchronized states of the coupled system of \eqref{dr}. In \cite{ambika} stability of the synchronized states of the environmentally coupled flows with no delay has been derived with some broad approximations; It was shown there no exact analysis is possible; nevertheless, the authors arrived at a condition that  predicts the stable complete (anti-) synchronized zone in parameter space. In the present case the scenario is more complex owing to the presence of delay. Let us start by considering $\psi$, $\theta$ and $\phi$ be the deviations from the synchronized states of the system variables in the Eq. \eqref{dr}. Then linearizing the system along with these deviations leads to
\begin{subequations}\label{lin}
\begin{align}
\dot{\psi}&=-a\psi(t)+b_1f'(x_{\tau})\psi_{\tau}+\epsilon_1\beta_1\phi,\label{linx}\\
\dot{\theta}&=-a\theta(t)+b_2f'(y_{\tau})\theta_{\tau}+\epsilon_1\beta_2\phi,\label{liny}\\
\dot{\phi}&=-\kappa\phi-\frac{\epsilon_2}{2}(\beta_1\psi+\beta_2\theta).\label{linz}
\end{align}
\end{subequations}
where, $u_{\tau}=u(t-\tau)$, $u=\psi, \theta$.

It is not possible to carry out an exact analysis of \eqref{lin}. To make the analysis possible we impose some constraints to it and consider this one as a special case. Let us consider the complete synchronized state of the systems, i.e, $x=y$ and hence, $x_{\tau}=y_{\tau}$. Also we define a new variable relating $\psi$ and $\theta$ in the following way:
\begin{equation}
\chi=\beta_1\psi+\beta_2\theta.\label{chi}
\end{equation}
With this, Eq. \eqref{lin} can be reduced to following equations (with $b_1=b_1=b$):
\begin{subequations}\label{td}
\begin{align}
\dot{\chi}  &=  -a\chi+ bf'(x_{\tau})\chi_{\tau}+\epsilon_1(\beta_1^2+\beta_2^2)\phi,\label{chid}\\
\dot{\phi}  &=  -\kappa\phi-\frac{\epsilon_2}{2}\chi.\label{phid}
\end{align}
\end{subequations}
Equation \eqref{td} also can not be solved in closed form. We can make further approach by considering $f'(x_{\tau})=\delta^\prime$, where $\delta^\prime$ is a constant; this approximation was also used and justified in \cite{ambika}. Define $\delta=b\delta^\prime$, and $\beta_1^2+\beta_2^2=2$, from \eqref{td} we get
\begin{subequations}\label{td1}
\begin{align}
\dot{\chi}  &=  -a\chi+\delta\chi_{\tau}+2\epsilon_1\phi,\label{chidot}\\
\dot{\phi}  &=  -\kappa\phi-\frac{\epsilon_2}{2}\chi.\label{phidot}
\end{align}
\end{subequations}
The characteristic equation of \eqref{td1} is given by
\begin{equation}\label{cheq}
\mbox{det}\left(
\begin{array}{cc}
\lambda+a-\delta e^{-\lambda\tau}  &  ~~~~-2 \epsilon_1\\
\frac{\epsilon_2}{2}               &  ~~~~\lambda+\kappa
\end{array}
\right)=0,
\end{equation}
which gives on evaluation 
\begin{equation}
\begin{split}
\lambda^2+(a+\kappa-\delta e^{-\lambda\tau})\lambda &-\delta\kappa e^{-\lambda\tau}\\
                                                    &~~~~~~+(a\kappa+\epsilon_1\epsilon_2)=0.\label{lam}
\end{split}
\end{equation}
The eigenvalue of the characteristic equation \eqref{lam} may be real or imaginary. Let us consider $\lambda=\mu\pm i\nu$. The synchronization will become just oscillatory if the eigenvalue be a purely imaginary one. Thus for the limiting case, we consider $\mu=0$, and $\lambda=\pm i\nu$. Substitution of this in the above and a comparison between the real and the imaginary parts yields
\begin{eqnarray}
-b\delta^\prime\nu\sin\nu\tau-b\delta^\prime\kappa\cos\nu\tau  &=&  \nu^2-(a\kappa+\epsilon_1\epsilon_2),\label{re}\\
-b\delta^\prime\nu\cos\nu\tau+b\delta^\prime\kappa\sin\nu\tau  &=&  -(a+\kappa)\nu.\label{im}
\end{eqnarray}
Squaring and adding Eq. \eqref{re} and \eqref{im}, we get 
\begin{equation}
\begin{split}
\nu^4+(a^2+\kappa^2&-2\epsilon_1\epsilon_2-b^2{\delta^\prime}^2)\nu^2\\
                                      &+(a\kappa +\epsilon_1\epsilon_2)^2-b^2{\delta^\prime}^2\kappa^2=0.\label{nu4}
\end{split}
\end{equation}
Consider $(a^2+\kappa^2-2\epsilon_1\epsilon_2-b^2{\delta^\prime}^2)=\Theta$ and $(a\kappa +\epsilon_1\epsilon_2)^2-b^2{\delta^\prime}^2\kappa^2=\Lambda$. Thus Eq. \eqref{nu4} reduces to 
\begin{equation}
\nu^4+\Theta\nu^2+\Lambda=0.\label{nutl}
\end{equation}
Eq. \eqref{nutl} has the following solution:
\begin{equation}
\nu^2=\frac{-\Theta\pm\sqrt{\Theta^2-4\Lambda}}{2}.\label{nu2}
\end{equation}
From the above equation it may be stated that $\nu^2$ must be real and positive, otherwise there is no purely imaginary roots of Eq. \eqref{lam}, and for this one requires $\Theta^2>4\Lambda$ along with $\nu^2>0$, that results 
\begin{equation}
\epsilon_1\epsilon_2<\lvert\kappa(\delta-a)\rvert.\label{ep12}
\end{equation}
This condition leads to the oscillatory solution of the complete synchronization (CS) indicating the loss of CS. Thus, the threshold condition of CS is
\begin{equation}
\epsilon_{1cr}\epsilon_{2cr}=\lvert\kappa(\delta-a)\rvert.\label{stability}
\end{equation}
Here, $\epsilon_{1cr}$ and $\epsilon_{2cr}$ are only the lower limits of the coupling strengths where a stable complete synchronized state can be achieved. If we increase the coupling strength beyond the critical values, then the synchronized state will be prevailed. In the next section we numerically confirm this result with proper choice of the effective value of $\delta$. Also note that \eqref{stability} is equally valid for the anti-synchronized (AS) state, because in that case consideration of $y(t)=-x(t)$ and $\beta_1=\beta_2=1$ does not alter the form of \eqref{lin}. 
\section{Numerical Simulation}
\label{sec:5}
The system equation \eqref{dr} is simulated numerically using Runge--Kutta algorithm with step size $h=0.01$. Following initial functions have been used for all the numerical simulations: for the $x$-system $\phi_x(t)=1$, for the $y$-system $\phi_y(t)=0.9$, and for the environment, i.e., $z$-system, $z(0)=0.85$. Also, the following system design parameters are chosen throughout the numerical simulations: $a=1$, $n=2.2$, $m=1$, $l=10$, $\kappa=1$; also we choose $b_1=b_2=1$, and $\tau=3.6$ to ensure that both the systems be in the hyperchaotic region \cite{banerjee12}. Unless otherwise stated, we use $\epsilon_1=\epsilon_2=\epsilon$. In all the computations, and real time and phase plane diagrams, a large number of iterations have been excluded to allow the system to settle to the steady state.
\subsection{Lyapunov exponent spectrum}
\begin{figure}
 \includegraphics[width=.49\textwidth]{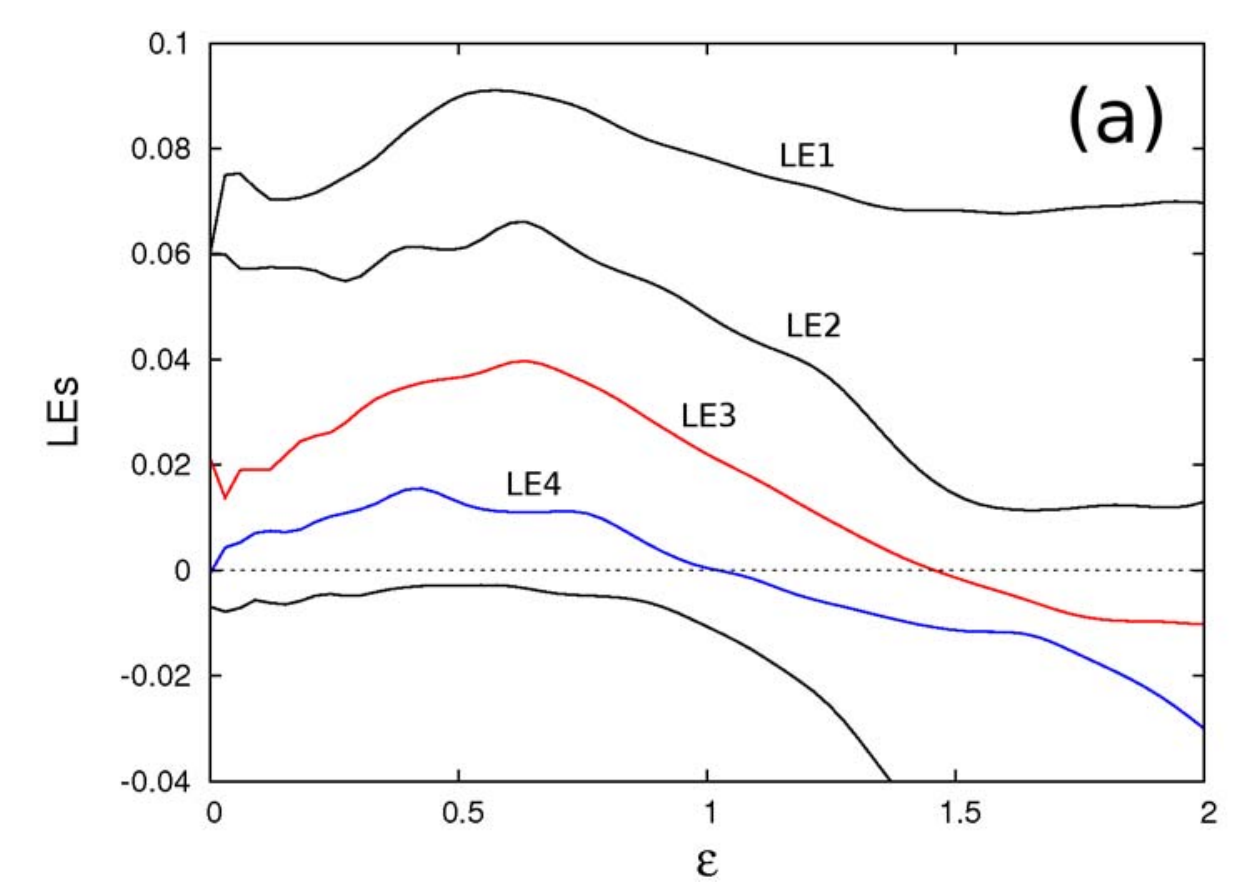}
\includegraphics[width=.49\textwidth]{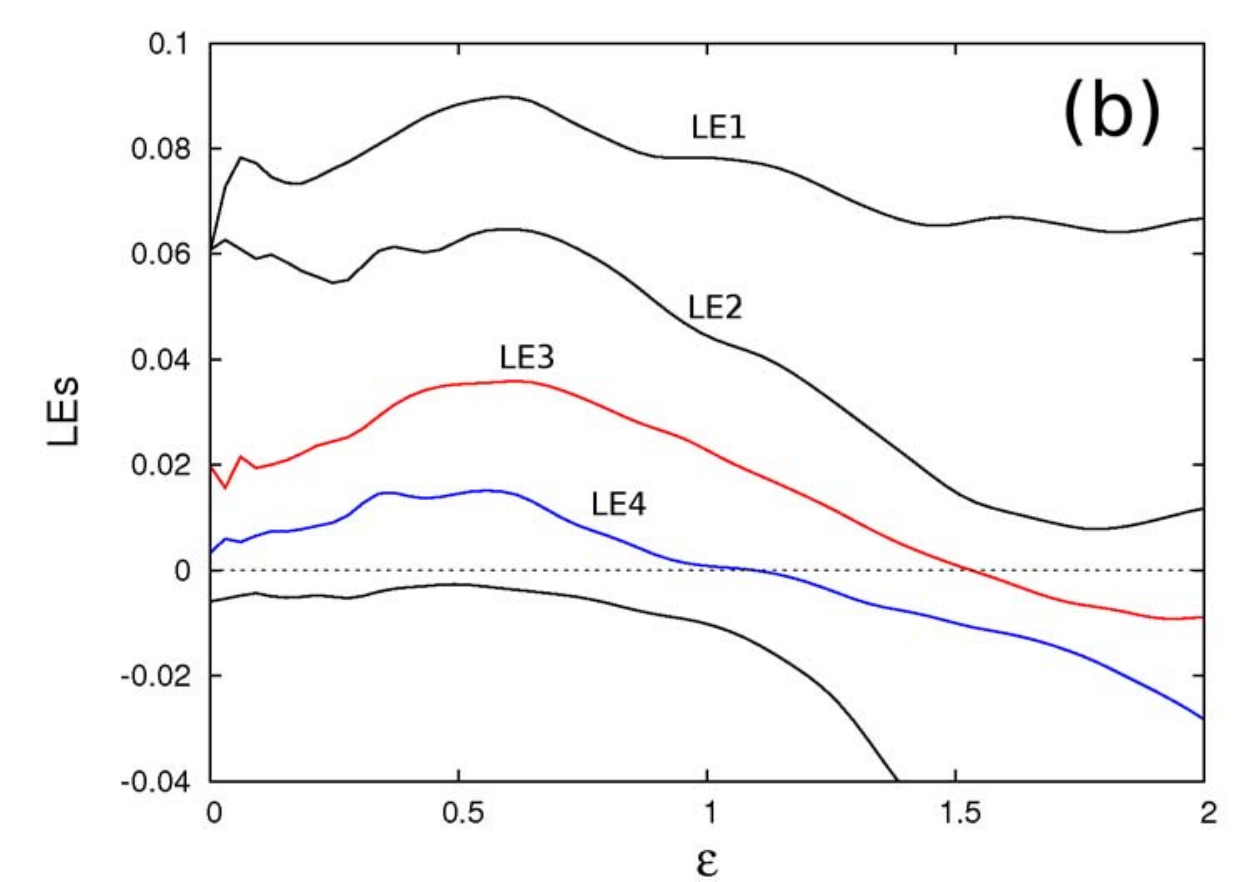}
 \caption{Lyapunov exponent spectrum of the coupled system. (a) $\beta_1=1, \beta_2=-1$. (b) $\beta_1=1, \beta_2=1$. Transitions of LE4 from positive to negative value through zero indicates phase synchronization, similar transition of LE3 indicates complete (anti-) synchronization.}
 \label{le}
\end{figure}

One can detect the onset of in-phase (inverse-phase) synchronization and complete (anti-) synchronization directly from the Lyapunov exponent (LE) spectrum of the coupled system \cite{piko}. We compute the LE spectrum in $\epsilon$--space ($\epsilon_1=\epsilon_2=\epsilon$) directly from \eqref{dr}. Among a large no of LEs we track the behavior of the first four LEs that are sufficient to detect the occurrence of phase or complete (anti-) synchronization with the variation of $\epsilon$ \cite{ambika}. For $\beta_1=1, \beta_2=-1$, Fig.~\ref{le}~(a) shows the first five LEs in $\epsilon$-space; it can be observed that with increase in $\epsilon$, the fourth largest LE (LE4) crosses the zero value to become negative at $\epsilon\approx1.1$ indicating the transition from unsynchronized state to in-phase synchronized state. With further increase in $\epsilon$, we observe that LE3 also becomes negative (from a positive value) at $\epsilon\approx1.48$, which indicates the transition from in-phase synchronization to complete synchronization. Similar case is shown in Fig.~\ref{le}~(b) for  $\beta_1=1, \beta_2=1$; here we can see that LE4 becomes negative at $\epsilon\approx1.2$ indicating inverse-phase synchronization and transition of LE3 from a positive value to a negative value at $\epsilon\approx1.56$ indicates anti-synchronization of the coupled systems.
\subsection{Time series and phase-plane plots}
\begin{figure}
\includegraphics[width=.49\textwidth]{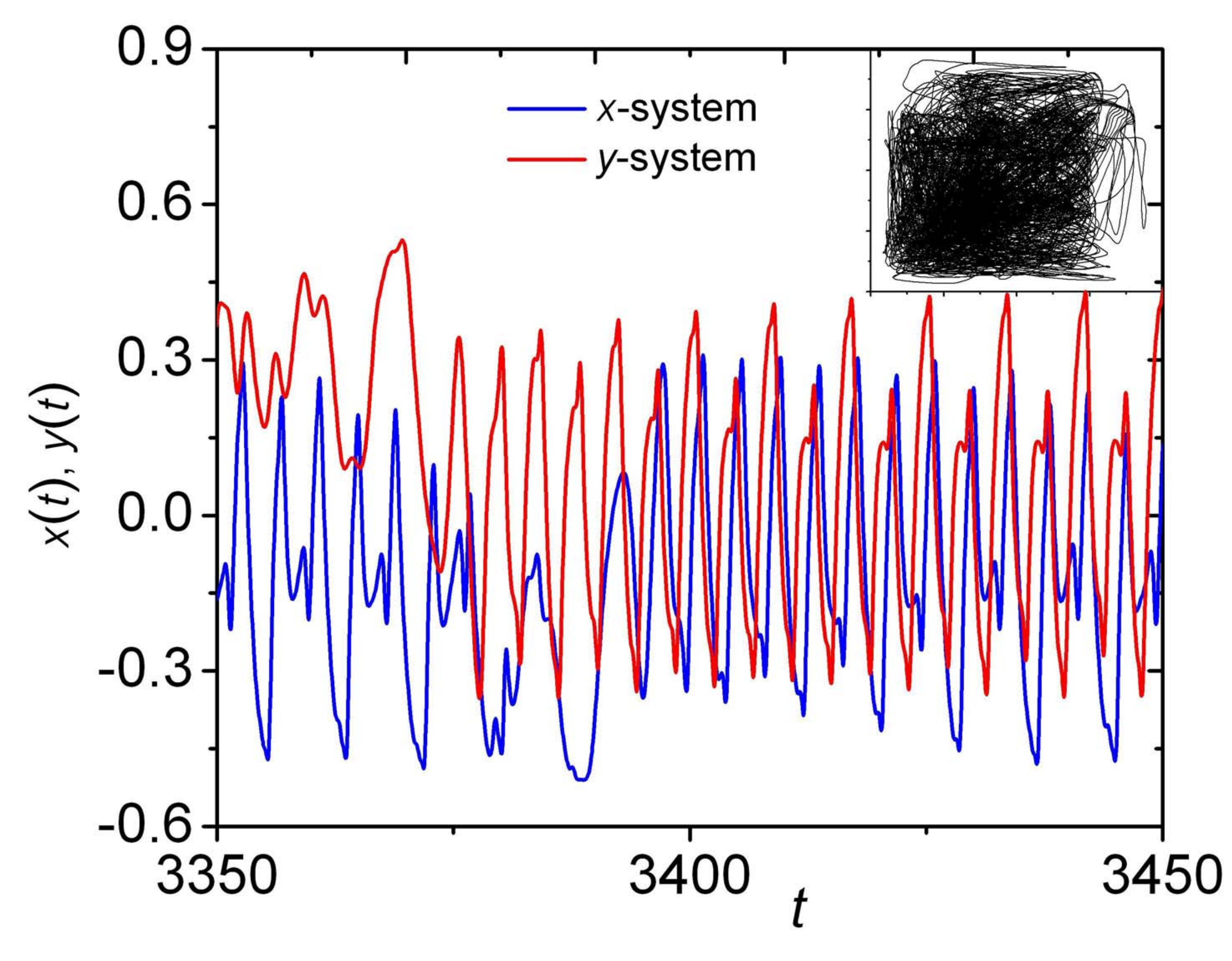}
\includegraphics[width=.49\textwidth]{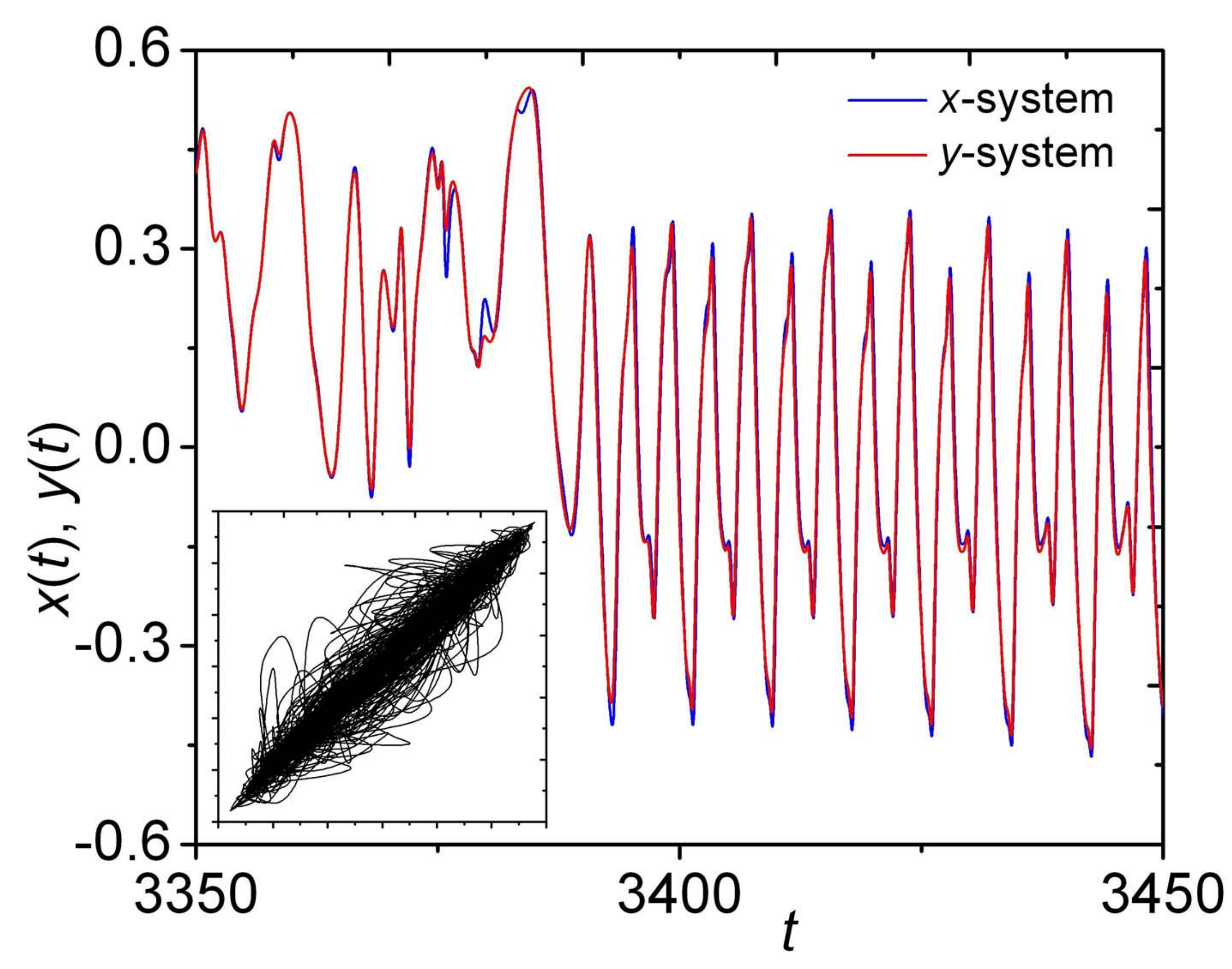}
\includegraphics[width=.49\textwidth]{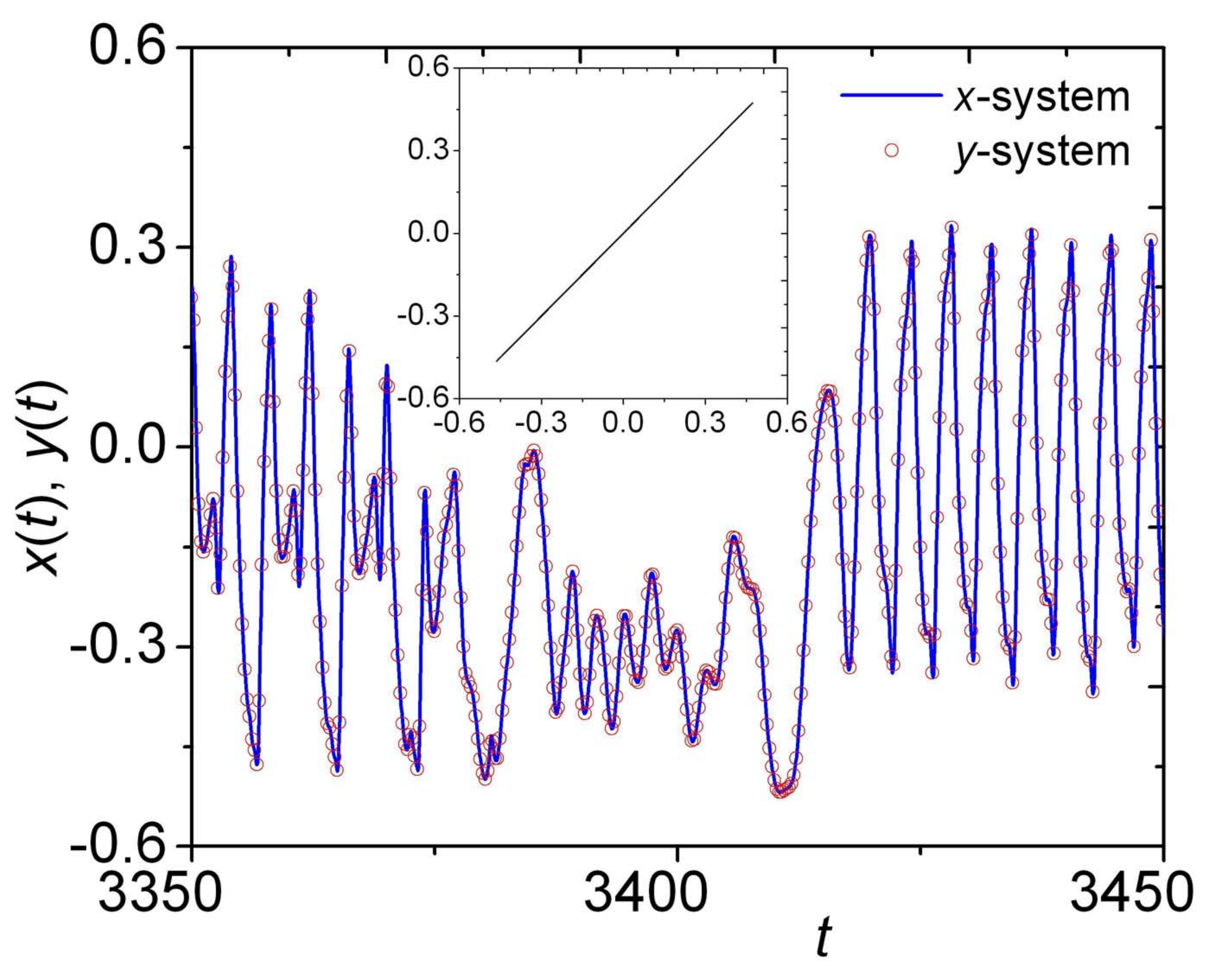}
\caption{$\beta_1=1$ and $\beta_2=-1$: Numerically obtained time-series of $x$ and $y$, and corresponding phase-plane plots in $x-y$ space. Upper panel: unsynchronized state ($\epsilon=0.2$), middle panel: in-phase synchronized state ($\epsilon=1.4$), lower panel: complete synchronized state ($\epsilon=1.6$).(Color figure online).}
\label{dts}
\end{figure}

\begin{figure}
\includegraphics[width=.49\textwidth]{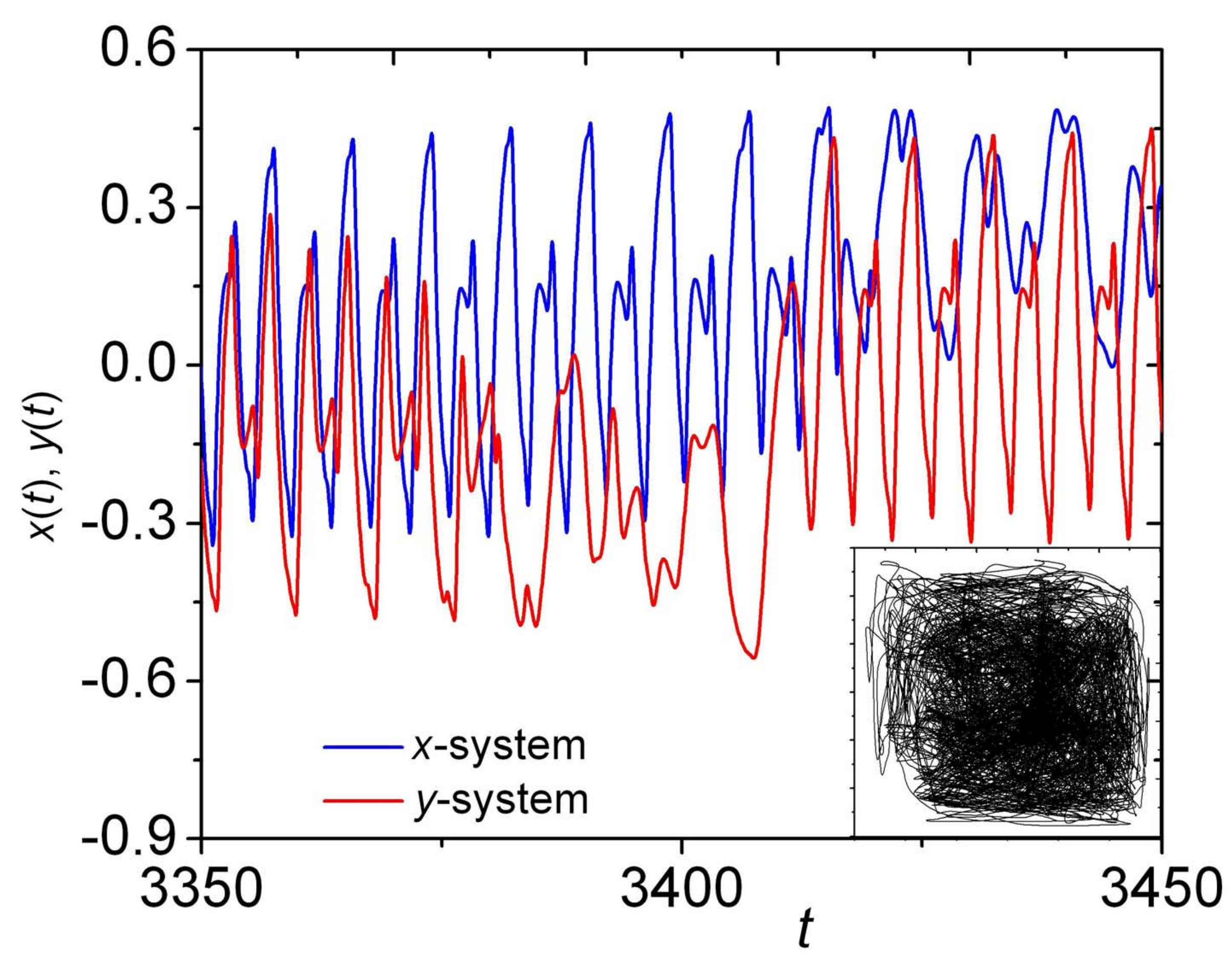}
\includegraphics[width=.49\textwidth]{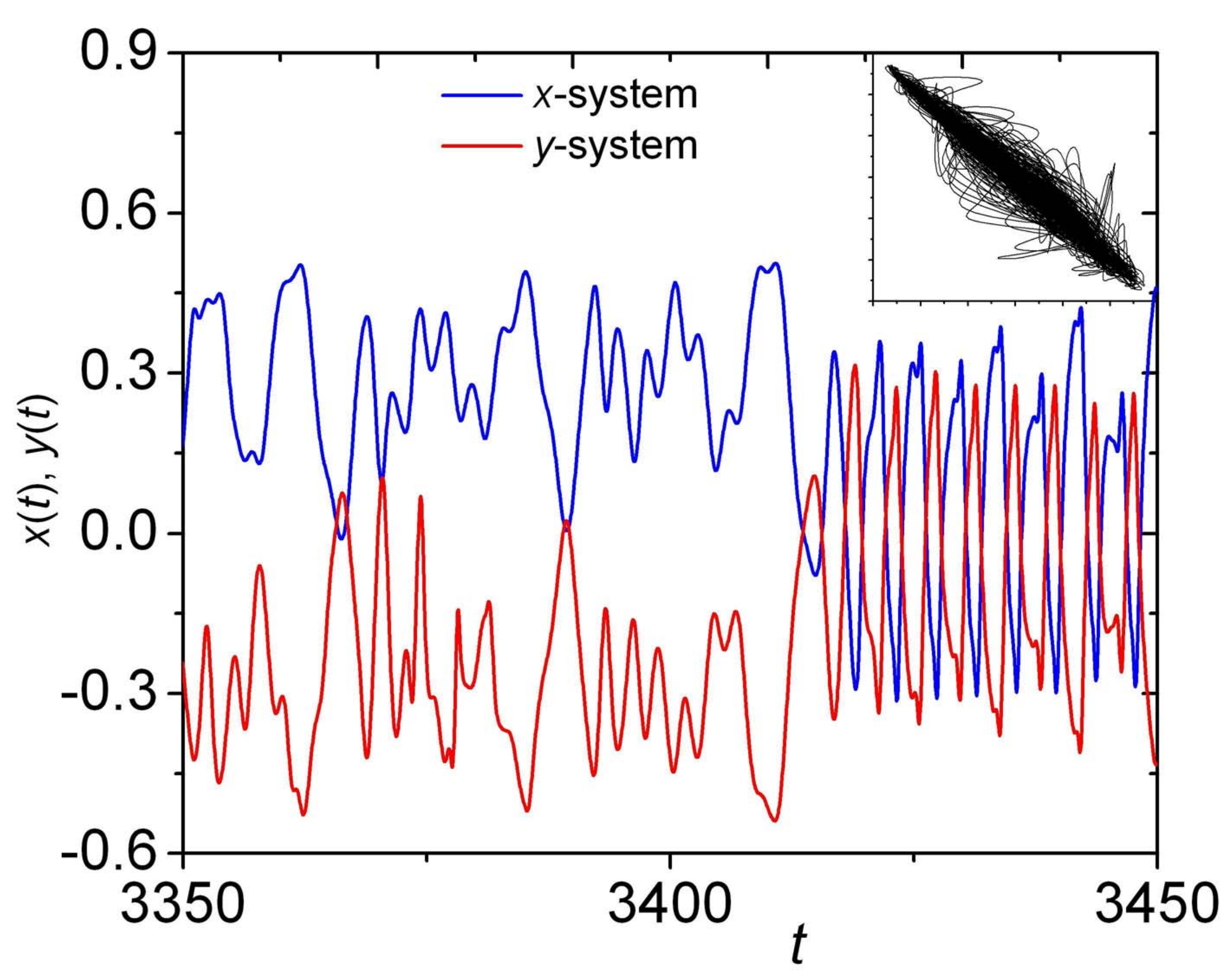}
\includegraphics[width=.49\textwidth]{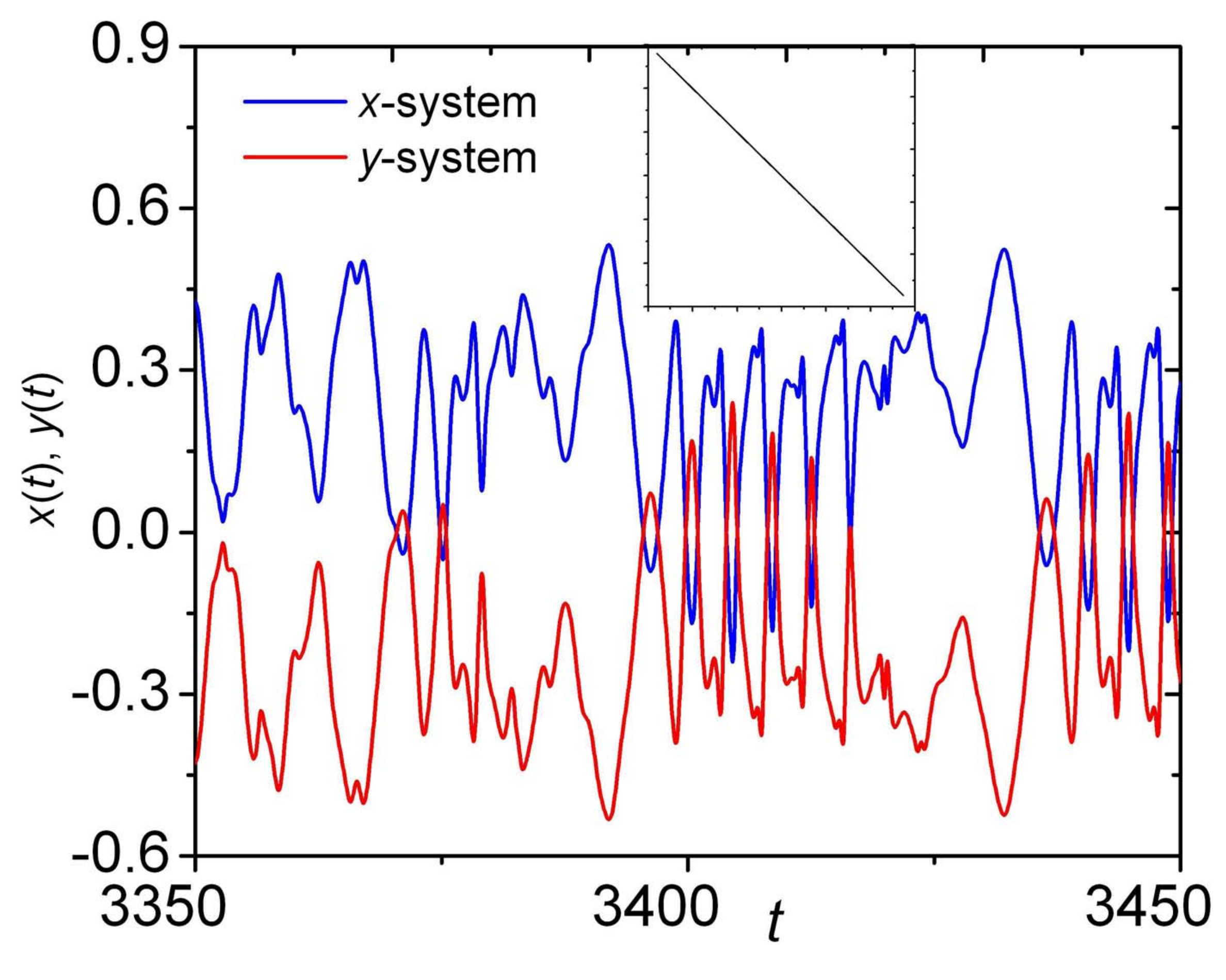}
\caption{$\beta_1=1$ and $\beta_2=1$: Numerically obtained time-series of $x$ and $y$, and corresponding phase-plane plots in $x-y$ space. Upper panel: unsynchronized state ($\epsilon=0.2$), middle panel: inverse-phase synchronized state ($\epsilon=1.4$), lower panel: anti-synchronized state ($\epsilon=1.6$).(Color figure online).}
\label{its}
\end{figure}

Guided by the above result, we choose three different coupling strengths, $\epsilon$ ($\epsilon_1=\epsilon_2=\epsilon$), and plot the time series and phase-plane plots for the illustrative examples of unsynchronized, phase synchronized and complete (anti-) synchronized states. Figure.~\ref{dts} shows this for $\beta_1=1, \beta_2=-1$. Upper panel of Fig.~\ref{dts} shows the unsynchronized case at $\epsilon=0.2$; corresponding phase-plane plot indicates no correlation between the $x$-system and $y$-system. Middle panel of Fig.~\ref{dts} shows the in-phase synchronized state at $\epsilon=1.4$; we can see from the corresponding phase-plane plot that although the phases of two systems are synchronized but their amplitudes still differ. Finally, lower panel of Fig.~\ref{dts} shows the case of complete synchronization at $\epsilon=1.6$; here both the systems become identical to each other.\\
Figure.~\ref{its} shows this scenario for $\beta_1=\beta_2=1$. At $\epsilon=0.2$ we can see the unsynchronized case (upper panel); at $\epsilon=1.4$ the system is inverse-phase synchronized (middle panel), and finally at $\epsilon=1.6$ we observe anti-synchronization (lower panel), where $x(t)$ and $y(t)$ have a $\pi$ phase shift and their amplitude levels are similar.
\subsection{Generalized autocorrelation function and CPR}
\begin{figure}
 \includegraphics[width=.49\textwidth]{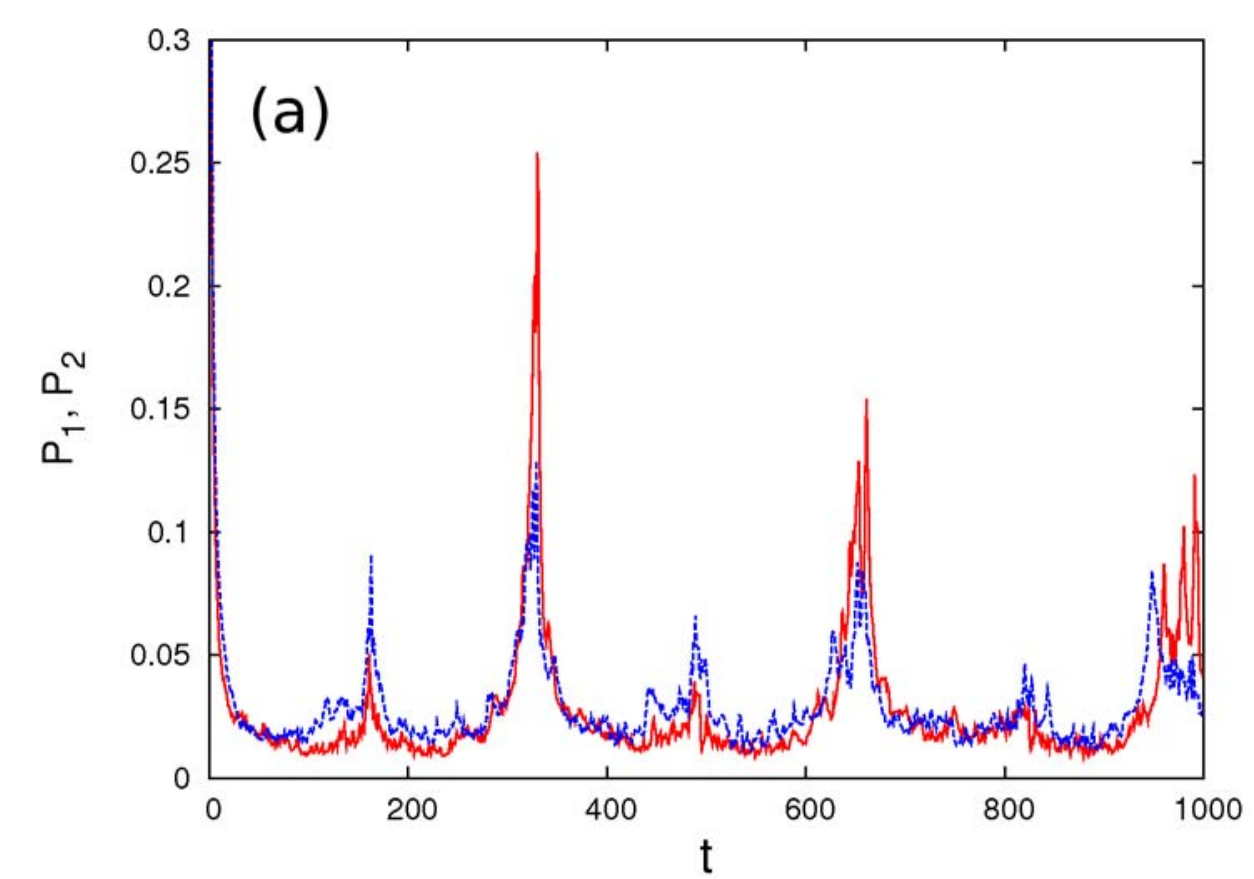}
\includegraphics[width=.49\textwidth]{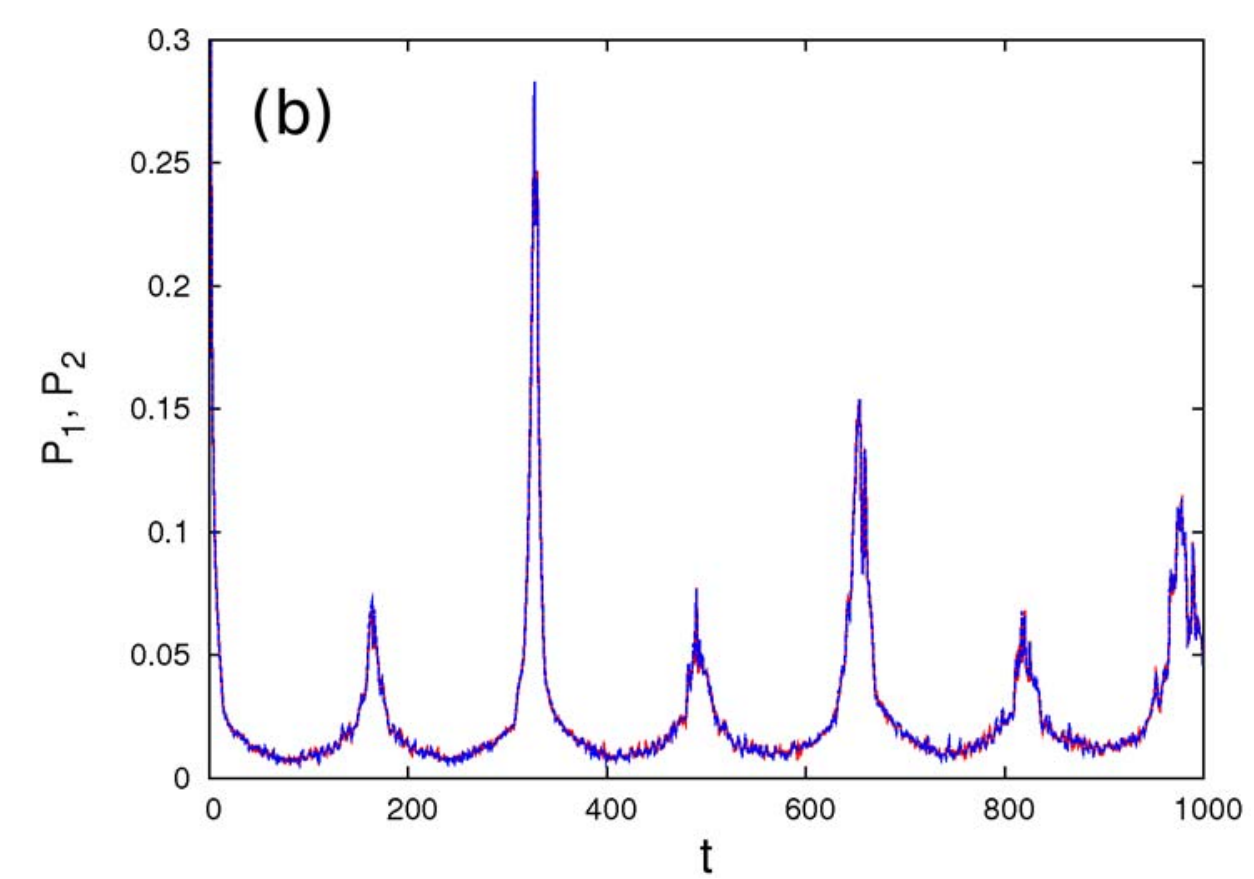}
 \caption{$\beta_1=1, \beta_2=-1$: Numerically obtained Generalized autocorrelation function, $P(t)$, of $x$-system ($P_1$, in red) and $y$-system ($P_2$, blue) in the unsynchronized state ($\epsilon=0.2$) (upper panel) and in-phase synchronized state ($\epsilon=1.4$) (lower panel). Notice the matching (difference) of the peaks of $P_1$ and $P_2$ in the in-phase synchronized (unsynchronized) case (for other parameter values see text).}
 \label{dcpr}
\end{figure}

\begin{figure}
 \includegraphics[width=.49\textwidth]{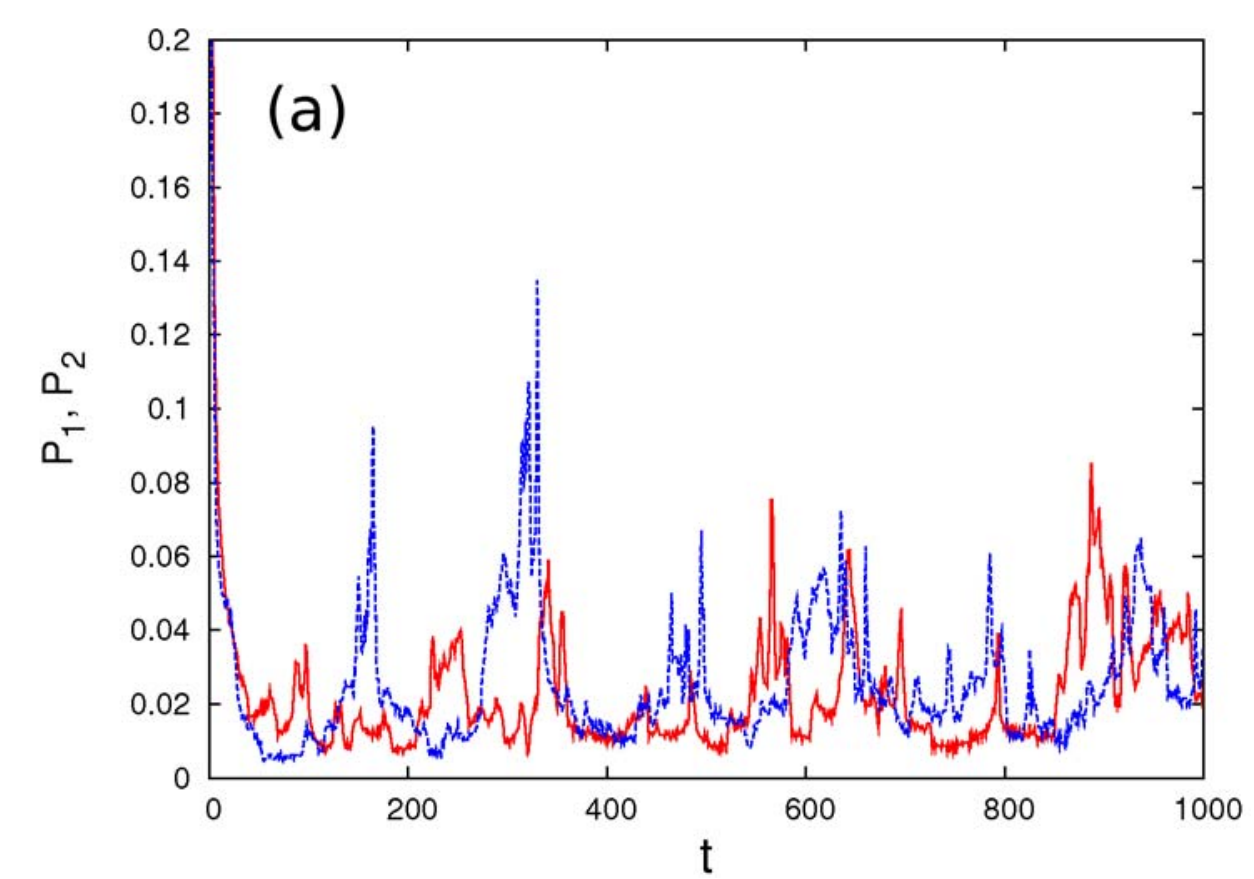}
\includegraphics[width=.49\textwidth]{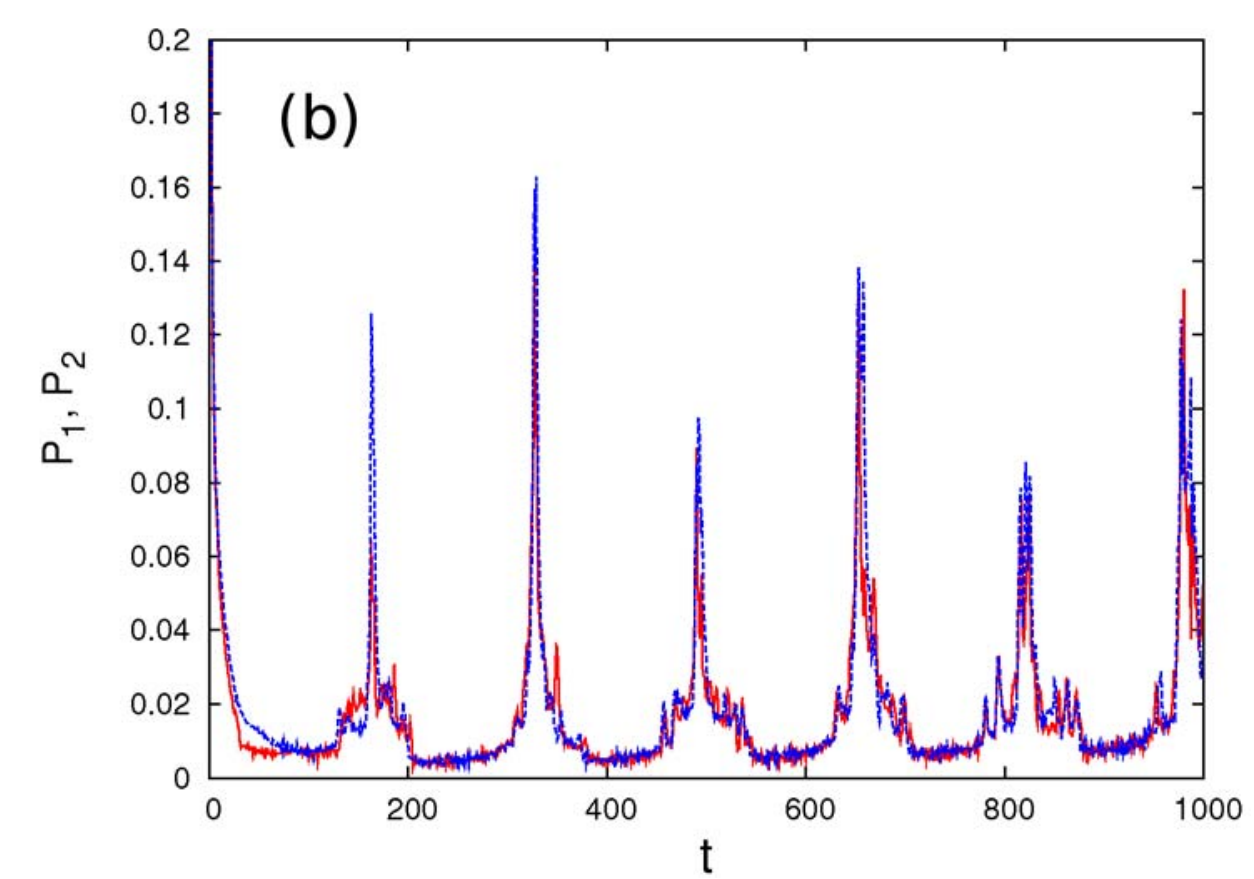}
 \caption{$\beta_1=1, \beta_2=1$: Numerically obtained Generalized autocorrelation function, $P(t)$, of $x$-system ($P_1$, in red) and $y$-system ($P_2$, in blue) in the unsynchronized state ($\epsilon=0.2$) (upper panel) and inverse-phase synchronized state ($\epsilon=1.4$) (lower panel). Notice the matching (difference) of the peaks of $P_1$ and $P_2$ in the inverse-phase synchronized (unsynchronized) case (for other parameter values see text).}
 \label{incpr}
\end{figure}

As discussed in Sect.\ref{sub:cpr} we compute $P(t)$ and CPR from numerical simulations. We use $N=5000$ and plot $P_{1,2}(t)$ for the unsynchronized and phase synchronized cases. Figure.\ref{dcpr} (a) and Fig.\ref{incpr} (a)  shows this at $\epsilon_1=\epsilon_2=0.2$ for ($\beta_1=1, \beta_2=-1$) and ($\beta_1=\beta_2=1$), respectively. The unmatched dominant peaks of $P_{1,2}(t)$ indicates the lack of phase synchronization in the coupled systems at this small value of coupling strength. We plot the same at $\epsilon_1=\epsilon_2=1.4$ (Figure.\ref{dcpr} (b) and Fig.\ref{incpr} (b)). Here the perfect matching of dominant peaks of $P_{1,2}(t)$ indicates the occurrence of PS; at this point we computed CPR, which is equal to 0.99 for  $\beta_1=1, \beta_2=-1$, and 0.987 for $\beta_1=\beta_2=1$ that also confirms the occurrence of PS in the coupled system.
\subsection{Concept of localized set}
\begin{figure}
 \includegraphics[width=.49\textwidth]{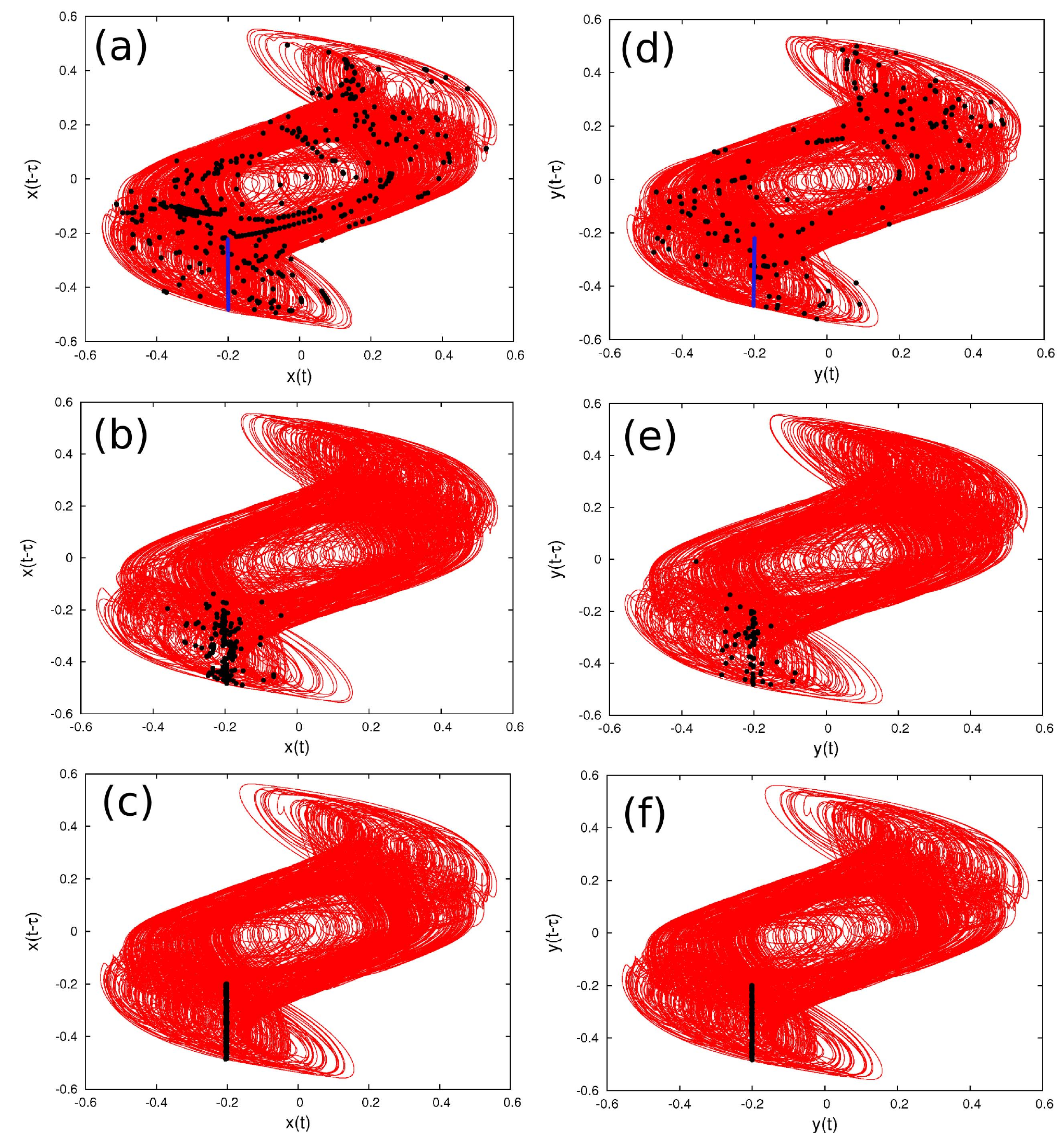}
 \caption{$\beta_1=1$ and $\beta_2=-1$: Numerically computed  Concept of localized sets: (a), (b), and (c): plot of $x(t)-x(t-\tau)$ along with the set $D$ (black dots) obtained by defining the following event in the $y$-system: $y(t)=-0.2$ and $y(t-\tau)\le-0.2$ (blue line in (d)). (d), (e), and (f): plot of $y(t)-y(t-\tau)$ along with the set $D$ (black dots) obtained by defining the following event in the $x$-system: $x(t)=-0.2$ and $x(t-\tau)\le-0.2$ (blue line in (a)). (a) and (d) represents unsynchronized state with $\epsilon=0.2$ (note that the black dots are scattered all around the attractors representing   that the attractors are incoherent). (b) and (e) represents phase synchronized states with $\epsilon=1.4$. (c) and (f) represents complete synchronized case with $\epsilon=1.6$. Note that in the last two cases black dots form a localized set that lives in a narrow region of the attractor.(Color figure online).}
 \label{dset}
\end{figure}

\begin{figure}
 \includegraphics[width=.49\textwidth]{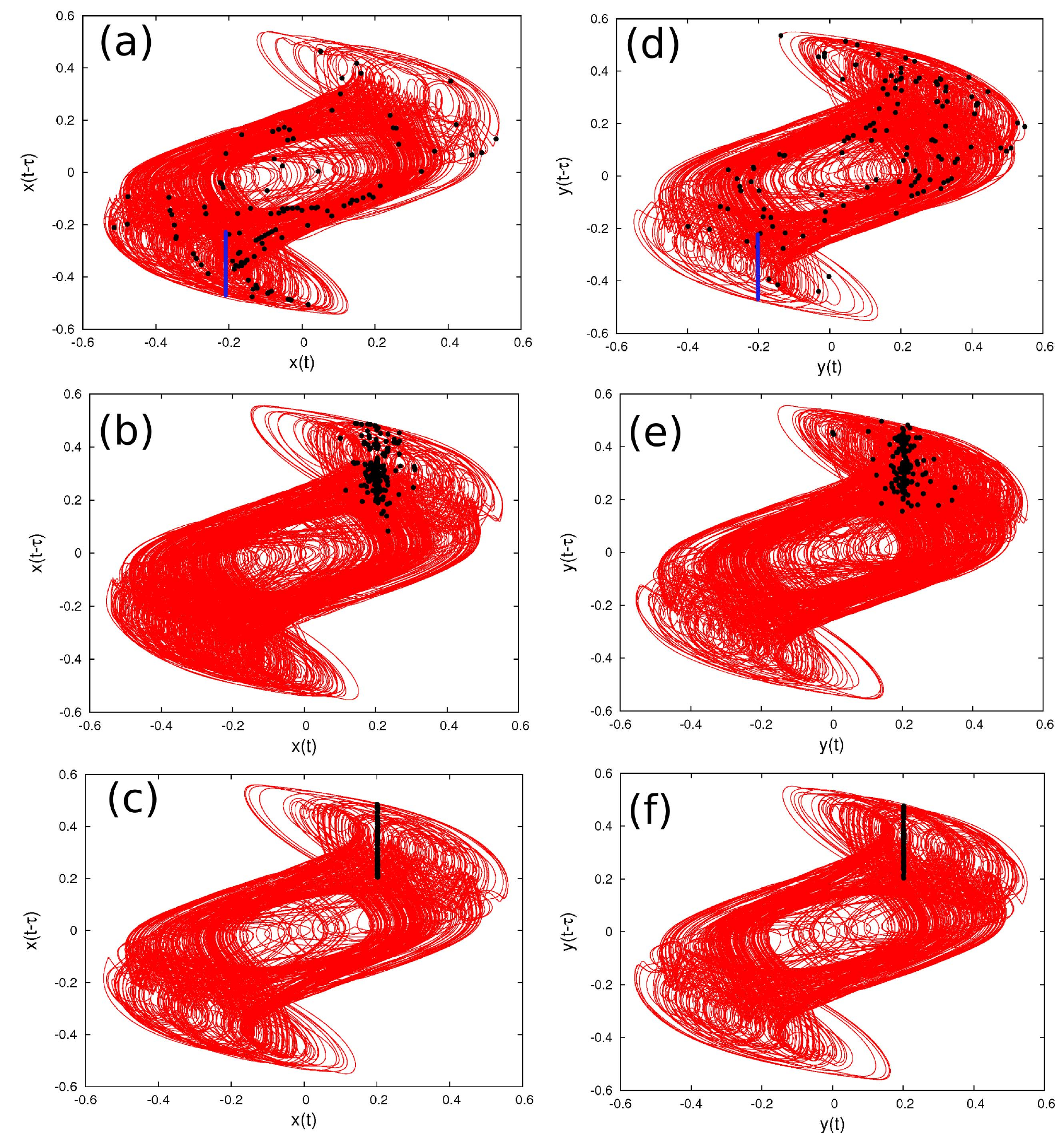}
 \caption{$\beta_1=1$ and $\beta_2=1$: Numerically computed  Concept of localized sets: (a), (b), and (c): plot of $x(t)-x(t-\tau)$ along with the set $D$ (black dots) obtained by defining the following event in the $y$-system: $y(t)=-0.2$ and $y(t-\tau)\le-0.2$ (blue line in (d)). (d), (e), and (f): plot of $y(t)-y(t-\tau)$ along with the set $D$ (black dots) obtained by defining the following event in the $x$-system: $x(t)=-0.2$ and $x(t-\tau)\le-0.2$ (blue line in (a)). (a) and (d) represents unsynchronized state with $\epsilon=0.2$ (note that the black dots are scattered all around the attractors representing   that the attractors are incoherent). (b) and (e) represents phase synchronized states with $\epsilon=1.4$. (c) and (f) represents anti-synchronized case with $\epsilon=1.6$.(Color figure online).}
 \label{inset}
\end{figure}

Further, we use the technique of concept of localized set to numerically confirm PS in the coupled system. Here, we define the following event in the $y$-system: $y(t)=-0.2$ and $y(t-\tau)\le-0.2$; whenever this event occurs in time, we track the values of the $x$-system and get a set of values of $x(t)$. Then we plot this set in the $x(t)-x(t-\tau)$ space. The similar process is repeated for the $y$-system with the following event in the $x$-system: $x(t)=-0.2$ and $x(t-\tau)\le-0.2$. Figure.~\ref{dset} shows this for  $\beta_1=1, \beta_2=-1$ and Fig.\ref{inset} represent the case for $\beta_1=\beta_2=1$ for various coupling strengths. Figure.~\ref{dset} (a,d) and Fig.~\ref{inset} (a,d) show the spreading of data set over the whole phase space for $\epsilon_1=\epsilon_2=0.2$ indicating phase-incoherent behavior of the coupled systems. At $\epsilon_1=\epsilon_2=1.4$ the data set forms a localized set in the phase space indicating the occurrence of phase synchronization (Fig.~\ref{dset} (b,e) and Fig.~\ref{inset} (b,e)). Finally, at $\epsilon_1=\epsilon_2=1.6$ we can see that Fig.~\ref{dset} (c,f) shows that the localized set becomes identical with the defined event itself for $\beta_1=1, \beta_2=-1$, indicating a complete phase synchronization. Whereas, for $\beta_1=\beta_2=1$, at $\epsilon_1=\epsilon_2=1.6$ Fig.~\ref{inset} (c,f) shows that the localized set is situated in a narrow region that is just opposite to the defined event, indicating anti-phase synchronization.
\subsection{Stability of synchronization in parameter space}
\begin{figure}
 \includegraphics[width=.49\textwidth]{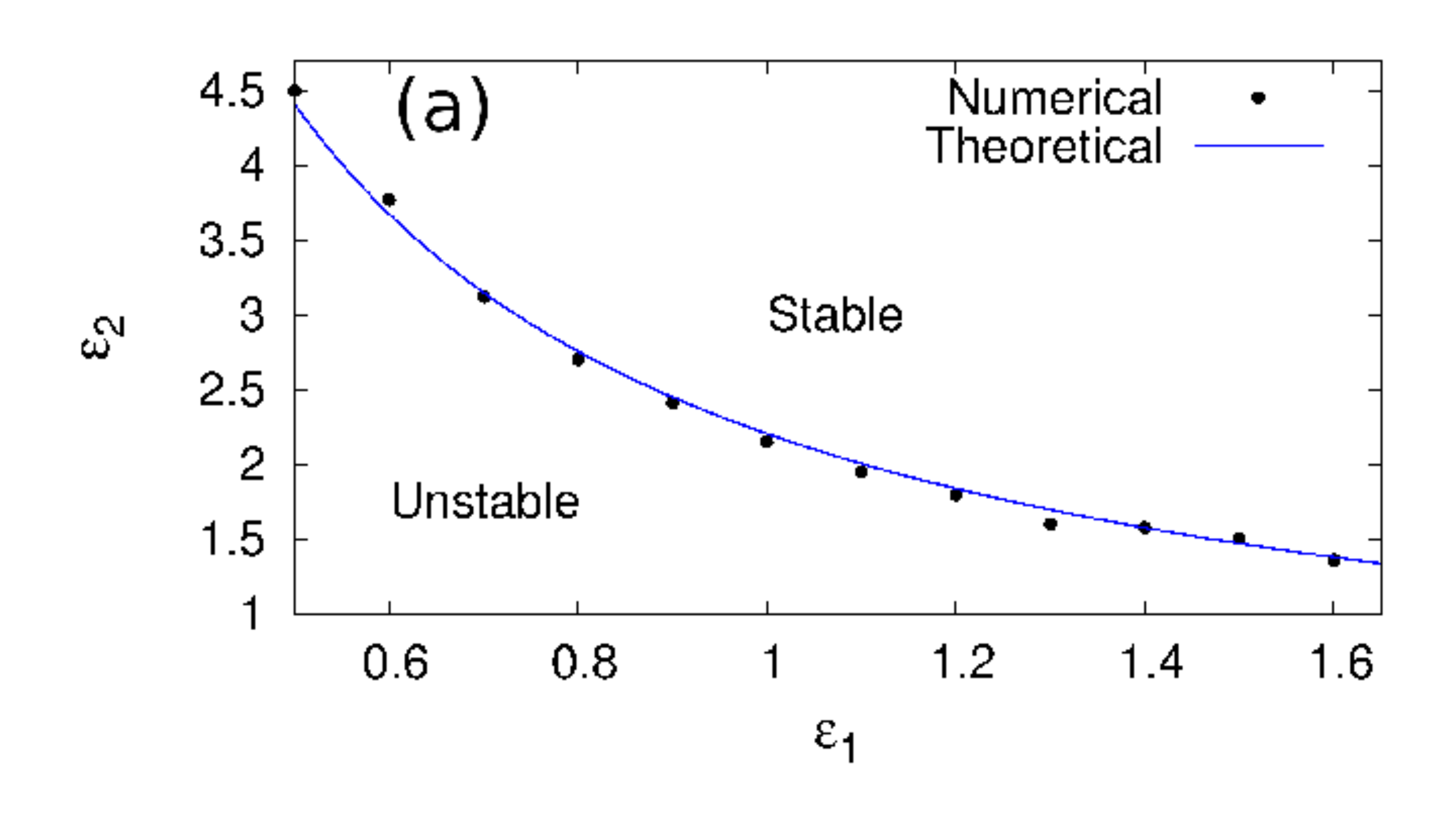}
\includegraphics[width=.49\textwidth]{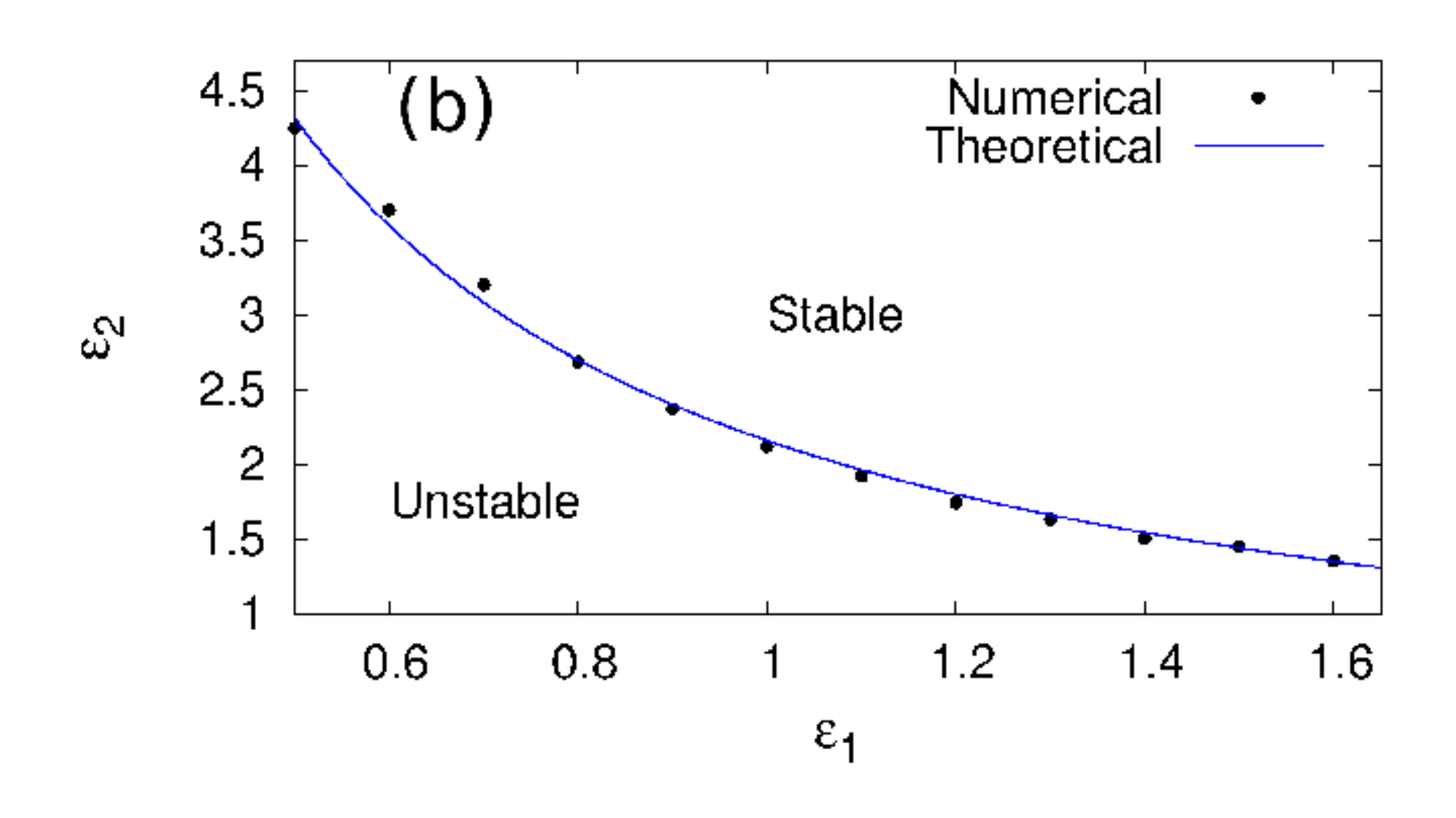}
 \caption{Stability curve in $\epsilon_1-\epsilon_2$ parameter space (a) $\beta_1=1$ and $\beta_2=-1$, (b) $\beta_1=1$ and $\beta_2=1$.}
 \label{e1e2}
\end{figure}

To support the theoretical result obtained analytically in \eqref{stability}, numerically we study the synchronization scenario in the $\epsilon_1-\epsilon_2$ space. We use all the above mentioned measures to detect the complete-synchronization (CS) and anti-synchronization (AS). The threshold value of $\epsilon_1$ and $\epsilon_2$ for which CS and AS occurs are plotted in Fig.\ref{e1e2} (a) and (b), respectively. The solid line represents the theoretical prediction and points represent the numerically obtained threshold values for which CS or AS just start. In theoretical curve, for CS, we use $\delta^\prime=3.20$ (also $a=1$, $b$=1, and $\kappa=1$); this value of $\delta^\prime$ is obtained for $x_{\tau}=\pm 0.0825$ (remember that $\delta^\prime=f'(x_{\tau})$). Similarly for AS we use $\delta^\prime=3.15$, that implies $x_{\tau}\approx\pm0.0832$. In both the cases the values of $x_{\tau}$, for which theoretical and numerical results match, lie well within the actual phase space where the system dynamics live. Thus the stability analysis agrees with the numerical results with some effective choice of $\delta^\prime$. Since the stability analysis relies on some broad approximations it requires further refinements and scopes are there to progress in that direction.
\section{Conclusion}
\label{sec:6}
In this paper we have reported different synchronization scenario of hyperchaotic time-delayed  systems coupled indirectly through a common environment. The system we have chosen is a first-order, nonlinear time-delayed system that posses a closed form mathematical function for the nonlinearity, shows hyperchaos even at a moderate or small time delay, and is convenient for the electronic circuit design. For the first time we have explored the experimental aspects of environmentally coupled time-delay systems. We have confirmed the occurrence of phase synchronization and complete (anti-) synchronization in the experimental circuit by using recurrence analysis and the concept of localized sets computed directly from the experimental time-series data. We have shown that with the proper choice of coupling parameters (i.e. $\beta_1=1, \beta_2=-1$), increase in the coupling strength results in a complete synchronized state from unsynchronized states via phase synchronized states. Also, for  $\beta_1=\beta_2=1$, with the increasing coupling strength we have observed a transition from unsynchronized states to the anti-synchronized state via inverse-phase synchronized states. To corroborate the experimental results we have presented a linear stability analysis of the complete (anti-) synchronized state, and also we perform detailed numerical simulations. Numerically we computed the Lyapunov exponent spectrum of the coupled system from where we have identified the zone of phase synchronization and complete (anti-) synchronization in the parameter space. We have used the recurrence analysis and the concept of localized set to numerically reconfirm the occurrence of phase synchronization in the system. Finally, using all the measures we have identified the parameter zone in the two-parameter space for the stable complete and anti synchronization, which agree well with the theoretical results.

The present study can be extended to explore the synchronization scenario of other environmentally coupled time-delay systems, e.g. Mackey-Glass system, Ikeda system, etc. One of the main features of the environmental coupling scheme is that here the system dynamics in the synchronized states and unsynchronized states have almost the same structure in phase-space \cite{ambika}; this particular feature makes this coupling scheme advantageous over other time-delay coupling schemes. From the academic interest, as the environmental coupling is very much relevant to biological systems, and delay is inherent in biological processes, thus the present study will be important for the study of synchronizations in the biological systems with delay. Apart from the academic interest, the present study is important from application point of view also; for example, since with environmental coupling one can make a transition from complete synchronization to anti-synchronization by simply changing the sign of $\beta_2$, thus this will be useful in easy implementation of hyperchaotic binary-phase-shift-keying (BPSK)-based digital communication systems. 

\section*{Acknowledgement}
Authors are grateful to Prof. B.C. Sarkar, Dept. of Physics, University of Burdwan, for the useful discussions and suggestions. One of the authors (D.B.) thankfully acknowledges the financial support provided by the University of Burdwan, Burdwan, India.

\end{document}